\documentclass[journal]{IEEEtran}

\usepackage{graphicx}
\usepackage{booktabs}
\usepackage{subfigure}
\usepackage{overpic}
\usepackage{amsmath}
\usepackage{flushend}
\usepackage{amssymb}
\usepackage{multirow}
\usepackage{makecell}
\usepackage{array}
\usepackage{stfloats}
\usepackage{algorithm,amsfonts}
\usepackage{epsfig, hyperref}
\usepackage{soul}
\usepackage{color, xcolor}
\usepackage{cite}

\hypersetup{
	colorlinks=true,
	linkcolor=cyan,
	urlcolor=red,
	citecolor=cyan,
}

\begin{document}

\title{Multi-scale Adaptive Fusion Network for Hyperspectral Image Denoising}
\author{Haodong Pan, Feng Gao, Junyu Dong, Qian Du
\thanks{This work was supported in part by the National Key Research and Development Program of China under Grant 2018AAA0100602, in part by the National Natural Science Foundation of China under Grant U1706218 and Grant 61871335. \emph{(Corresponding author: Feng Gao)}

Haodong Pan, Feng Gao, and Junyu Dong are with the School of Computer Science and Technology, Ocean University of China, Qingdao 266100, China.

Qian Du is with the Department of Electrical and Computer Engineering, Mississippi State University, Starkville, MS 39762 USA.}
}

\markboth{IEEE Journal of Selected Topics in Applied Earth Observations and Remote Sensing}%
{Shell}

\maketitle

\begin{abstract}

Removing the noise and improving the visual quality of hyperspectral images (HSIs) is challenging in academia and industry. Great efforts have been made to leverage local, global or spectral context information for HSI denoising. However, existing methods still have limitations in feature interaction exploitation among multiple scales and rich spectral structure preservation. In view of this, we propose a novel solution to investigate the HSI denoising using a Multi-scale Adaptive Fusion Network (MAFNet), which can learn the complex nonlinear mapping between clean and noisy HSI. Two key components contribute to improving the hyperspectral image denoising: A progressively multiscale information aggregation network and a co-attention fusion module. Specifically, we first  generate a set of multiscale images and feed them into a coarse-fusion network to exploit the contextual  texture correlation. Thereafter, a fine fusion network is followed to exchange the information across the parallel multiscale subnetworks. Furthermore, we  design a co-attention fusion module to adaptively  emphasize informative features from different scales, and thereby enhance the discriminative learning capability for denoising. Extensive experiments on synthetic and real HSI datasets demonstrate that the proposed MAFNet has achieved better denoising performance than other state-of-the-art techniques.  Our codes are available at \verb'https://github.com/summitgao/MAFNet'.

\end{abstract}

\begin{IEEEkeywords}
Deep learning, Hyperspectral image denoising, fusion module, neural network, global contextual correlation.
\end{IEEEkeywords}

\IEEEpeerreviewmaketitle

\section{Introduction}

\IEEEPARstart{H}{yperspectral} images (HSIs) have developed rapidly with the maturity of remote sensing technology. HSIs have been extensively applied in land cover classification \cite{Bioucas13_grsm, zhang2021topological, zhang2022graph, hong2020graph}, semantic segmentation \cite{hong2020multimodal}, change detection \cite{tong13_jstars, marinelli19_tgrs, wang19_tgrs, zhan20_jstars, ou22jstars, wang22tgrs}, oil spill monitoring \cite{sun19_tgrs, kang22tgrsoil, li21jstarsoil}, and geographic transport prediction \cite{liu20_tgrs}. In these applications, high-quality images are commonly desired. However, during the HSI acquisition, some noise corruptions are inevitable and degrade the visual quality considerably \cite{lu13_tgrs}. Hence, removing noise from the acquired HSI is a critical step for many remote sensing applications \cite{jia20_grsl}. The task of highly efficient HSI denoising has recently captured numerous research attention.

The HSI denoising task aims to recover an underlying clean image $I$ from a noise observed data $I_N$. The degradation model is commonly formulated as $I_N = I + I^*_N$. Here, $I^*_N$ denotes the mixed noise. To solve the ill-posed inverse problem, many methods model the prior knowledge of the clean image $I$ to constrain the solution space. Total variation \cite{yuan12_tgrs, du18_ijrs, zhang20_tgrs, hu20_tgrs}, sparsity-driven models \cite{rasti20_grsl, chen20_tc, cao19_jstars} and low-rank representations \cite{lowrank1, lowrank2, lowrank3, lowrank4} are commonly used. The intrinsic structures in HSIs are modeled by these optimization-based models for noise removal.

\begin{figure}
\centering
\includegraphics [width=3.2in]{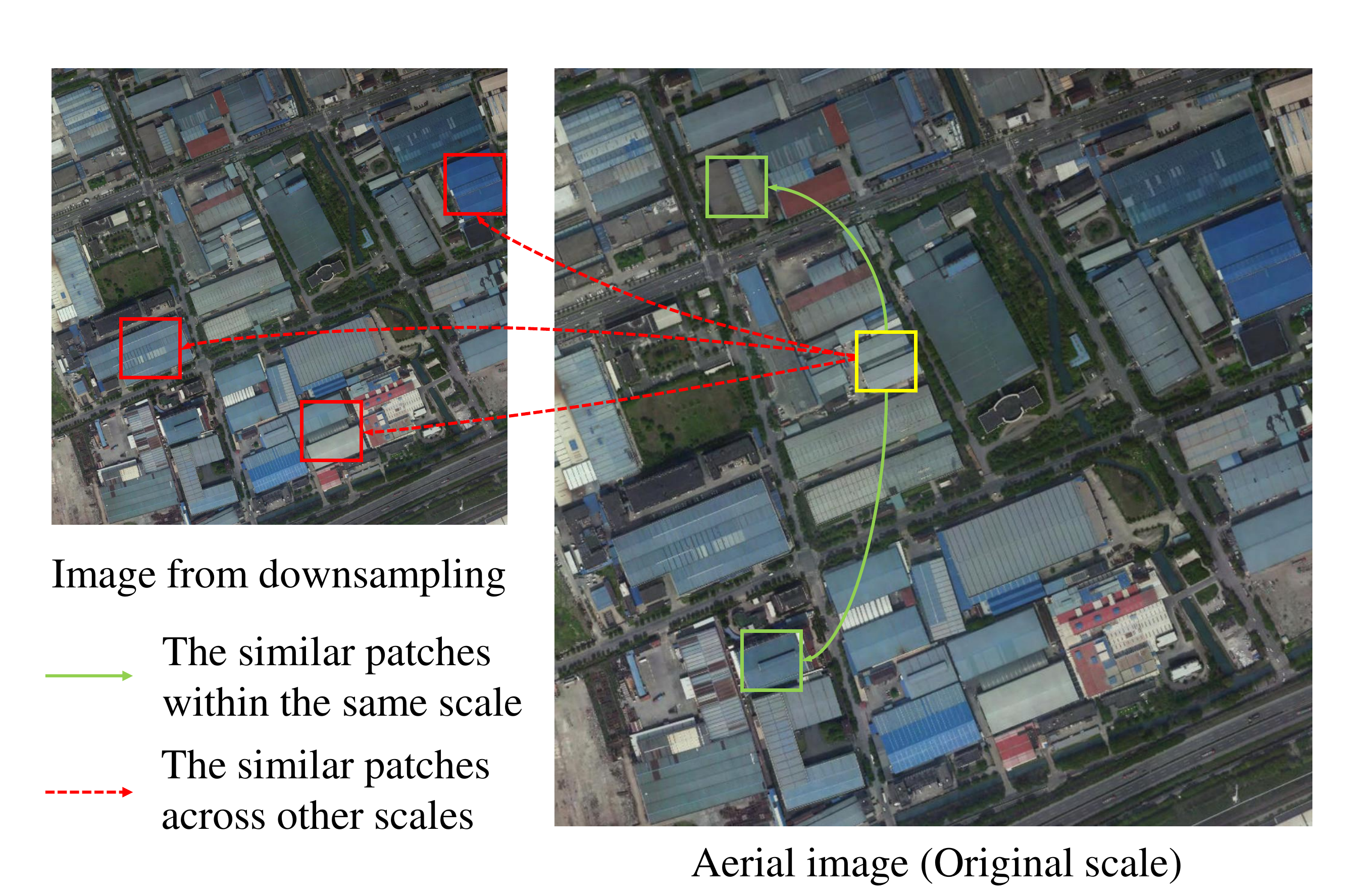}
\caption{Illustration of the multiscale representation of high resolution remote sensing image. Specially, there are similar patches in different scales, both in the same scale (highlighted in green) and across different scales (highlighted in red), can make contribution as additional information to reconstruct the target area (highlighted in blue).}
\label{fig_ms}
\end{figure}

During the past few years, the deep learning-based model for low-level vision tasks has demonstrated significant potential and performance improvement. It has been extensively employed to image restoration tasks such as compression artifact reduction \cite{chen21spl}, image denoising \cite{zou21tip, zhang21tip}, and image super-resolution \cite{yu20tgrs, xie21tgrs}. In HSI denoising, the deep learning-based model has yielded excellent results recently. Deep convolutional neural networks (CNNs) are capable of exploiting rich feature representations from large-scale training data, instead of hand-crafted features, which are designed according to prior knowledge.  Most existing CNN-based HSI denoising methods follow high-resolution feature processing \cite{chang19tgrs, liu19tgrs, yuan19tgrs}. These methods do not employ any downsampling operation; and hence, more accurate spatial details can be retained in the denoising results. However, the contextual information is inclined to get lost due to the limited receptive field. To effectively encode the contextual information, some researchers employ an encoder-decoder architecture \cite{dong19tci, maffei20tgrs}. The input HSI is progressively mapped into a low-resolution representation and then gradually mapped to the original resolution. The broad context can be learnt in the low-resolution representation \cite{hong2020joint}. However, some fine spatial details are hard to be reserved. Such detailed information is hard to be recovered in the decoding stages.

It is crucial to encode the contextual information and preserve the spatial details simultaneously for robust HSI denoising. However, it is a non-trivial task due to the following challenges: 1) \textbf{Tradeoff between the spectral-spatial detail preservation and contextual information modeling}. HSI denoising creates pixel-to-pixel correspondence from the observed data with complex noise to the clean image, and it is essential to preserve the detailed spectra and texture via high-resolution feature processing networks. However, the contextual information is hard to be encoded while preserving the spatial details by existing models. Hence, how to preserve the detailed spectra and texture while encoding the contextual information effectively is of great significance. 2) \textbf{Multiscale information aggregation}. Image contents from multiple scales encode complementary information for feature representation, as shown in Fig. \ref{fig_ms}. Existing multiscale models rarely exchange information across different scales flexibly. The correlations among different scales have not been fully exploited. Therefore, how to aggregate multiscale information into a unified framework is a tricky task. Wang et al. \cite{wang2020deep} proposed a deep High-Resolution representation Network (HRNet) for visual recognition. It maintains high-resolution representations in the whole network and repeatedly exchanges information among multi-resolution features. It has been widely used for human pose estimation \cite{hrformer21neurips}, semantic segmentation \cite{Gu_2022_CVPR, Wang_2021_ICCV, chu2021neurips}, and multispectral image classification \cite{li21igarss}. If such framework could be introduced into HSI restoration, the denoising performance could be further improved.

To solve the aforementioned issues, we proposed a \underline{M}ulti-scale \underline{A}daptive \underline{F}usion \underline{Net}work (MAFNet) for hyperspectral image denoising. The framework of MAFNet is illustrated in Fig. \ref{fig_framework}. Specifically, we first generate a set of multiscale images and feed them into a coarse-fusion network to exploit the contextual texture correlation. Meanwhile, a fine fusion network is followed to exchange information across the parallel multiscale subnetworks. Furthermore, we design a co-attention fusion module to adaptively emphasize informative features from different scales, and thereby enhance the discriminative learning capability for denoising. We next adopt a reconstruction loss together with a global gradient regularization to optimize the network. We conduct extensive experiments on five publicly available datasets. The experimental results show that the proposed MAFNet outperforms several state-of-the-art baselines. Our MAFNet differs from HRNet in two respects: First, in the coarse-fusion network, the information interaction direction is only from the low-resolution features to the high-resolution features. In this stage, we aim to increase the receptive field to capture more content. Second, in multiscale feature fusion, HRNet transforms features from different resolutions to the same size and then concatenates them together as the fusion output. The proposed MAFNet use adaptive instance and co-attention mechanism for adaptive feature fusion.

The contributions of this work can be summarized as follows:

\begin{itemize}
    \item We propose a novel hyperspectral image denoising model MAFNet, which 
    progressively fuses multiscale information. Hence, the global contextual information modeling and spatial detail preservation can be achieved simultaneously.
    
    \item We present a co-attention fusion module to dynamically select the useful features from each scale subnetwork, enhancing the discriminative learning capability. Thereby, multiscale information is adaptively aggregated, and the correlations among different scales are concurrently enhanced.
    \item Extensive experiments are conducted on two benchmark datasets, which demonstrates the rationality and effectiveness of the proposed MAFNet. Meanwhile, we have released our codes to benefit the remote sensing image restoration community.
\end{itemize}

The reminder of this paper is organized as follows: In Section II, we review closely related HSI denoising methods. The details of MAFNet are described in Section III. Experiments on several datasets on HSI datasets are presented in Section IV. Conclusions are drawn in Section V.

\section{Related Work}

The HSI denoising is an essential step to improve the image quality before interpretation. To date, a great number of methods have been proposed to reduce the noise in HSIs. In this paper, the existing methods are classified into three categories and introduced respectively.

\subsection{Filter-Based Methods for HSI Denoising}

In the beginning, the filter operator is generally used for HSI denoising, and it aims to separate the clean image from noisy signals by non-local means filter or Fourier transform. Othman et al. \cite{othman06tgrs} presented a wavelet shrinkage method for HSI denoising. The method benefits from the feature dissimilarity between the spectral and spatial domains. Zelinski et al. \cite{zelinski06igarss} proposed a method based on wavelet decomposition and sparse approximation for HSI denoising, thereby exploiting the correlation between bands and higher quality band information. Maggioni et al. \cite{maggioni13tip} proposed BM4D for volumetric data denoising, which is an extension of the BM3D filter. It embeds the grouping and collaborative filtering paradigms, thus integrating spatial and frequency domain filtering. Letexier and Bourennane \cite{letexier08tgrs} used the multidimensional Wiener filter for HSI denoising. Quadtree decomposition is also utilized to keep local characteristics. These filter-based methods are sensitive to the transform function and, therefore, can hardly remove the mixed noise in HSIs.

\subsection{Model-Based Methods for HSI Denoising}

The model-based method is the most popular representation tool for HSI denoising. Total variation \cite{yuan12_tgrs, du18_ijrs, zhang20_tgrs, hu20_tgrs}, sparsity-driven models \cite{rasti20_grsl, chen20_tc, cao19_jstars}, and low-rank representations \cite{lowrank1, lowrank2, lowrank3, lowrank4} are commonly employed to establish an optimization model for HSI denoising. For instance, Yuan et al. \cite{yuan12_tgrs} presented an adaptive total variation model for HSI denoising, which simultaneously models the spectral and spatial noise distribution. Zhang et al. \cite{zhang20_tgrs} proposed a HSI denoising method based on nonlocal low-rank tensor decomposition, in which the nonlocal similarity between the data cubes is captured to build a clean image. Zhuang et al. \cite{zhuang20jstars} proposed a denoising method that exploited the low-rank structure of the HSI data, utilizes the low-rank and self-similar characteristics contained in HSI for sparse and compact representation. Cao et al. \cite{cao19jstars} presented a subspace-based nonlocal low-rank method for HSI denoising.  These methods achieve satisfying performance due to the comprehensive consideration of the image's prior information.

\subsection{CNN-Based Methods for HSI Denoising}

Recently, research on natural image restoration has been dominated by the deep CNN in recent years \cite{chen21spl, zou21tip, zhang21tip, yu20tgrs, gu19iccv, liu20cvpr, kim20cvpr}. Chang et al. \cite{chang19tgrs} introduce the deep CNN model for HSI denoising, uses residual learning, dilated convolution and multi-channel filtering to enhance the ability to express spectral features. Liu et al. \cite{liu19tgrs} presented a 3-D atrous convolution method for HSI denoising. Atrous convolution was employed to enlarge the receptive fields. Zhang et al. \cite{zhang19tgrs} employed the gradient learning strategy to capture the intrinsic and deep features of HSI. Cao et al. \cite{cao21tgrs} proposed two global reasoning modules to exploit the contextual information along the channel and spatial dimensions, respectively. Both modules are combined in dense CNNs to exploit rich feature representations. Lin et al. \cite{lin20tip} built a CNN-constrained nonnegative matrix factorization model for HSI denoising,  realizes the optimization of noisy images through three stages: update of the spectral matrix, update of the abundance matrix, and estimation of the sparse noise. Wei et al. \cite{wei21tnnls} designed QRNN3D, the model adopts the encoder-decoder structure, realizes the use of spatial pixel correlation information and spectral global information by building a three-dimensional recurrent unit, and uses the special structure of alternating directions to eliminate unreasonable causal dependencies. Capable of flexible processing of hyperspectral images.

Leveraging the powerful linear modeling capability of the deep CNN, these methods achieved promising performance on HSI denoising. However, existing CNN-based methods rarely build feature communications between cascaded multiscale layers; thus, the correlated noise information across different scales is not fully exploited.

\section{Methodology} 

\begin{figure*}[ht]
\centering
\includegraphics [width=6.8in]{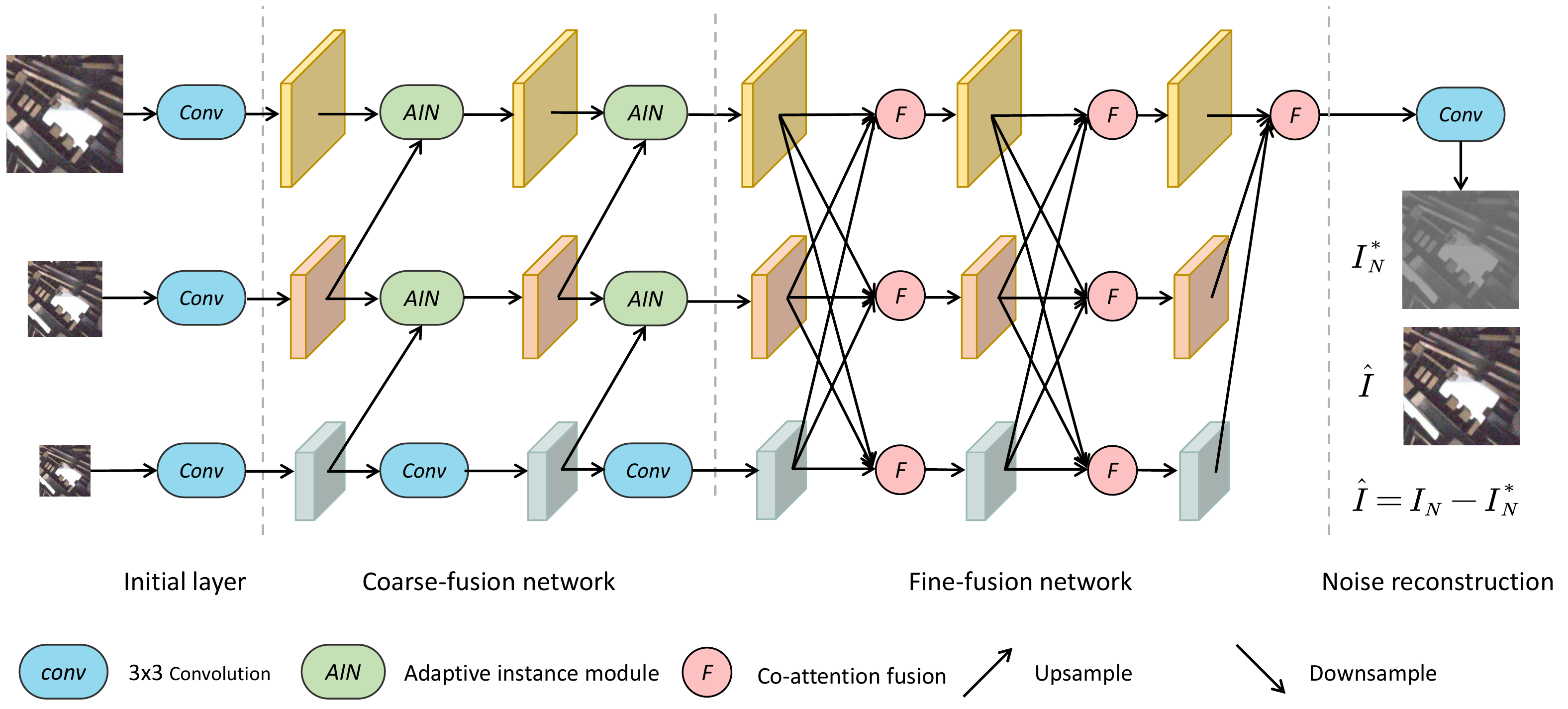}
\caption{Framework of the Multi-scale Adaptive Fusion Network (MAFNet). MAFNet is composed with initial layer, coarse-fusion network, fine-fusion network and noise reconstruction. Initial layer obtains the feature representation from the HSI image, then coarse-fusion network realizes the information transfer from the low-scale network to the high-scale network, then fine-fusion network fully integrates the contextual global information, and finally we reconstruct noise to get denoising images. For upsampling, we use transposed convolution to get the hight-resolution representation of the feature.}
\label{fig_framework}
\end{figure*}

In this section, we present the MAFNet for hyperspectral image denoising, which exploits the inherent correlation of noise across multiple scales. As illustrated in Fig. \ref{fig_framework}, MAFNet consists of four parts: initial layer, coarse-fusion network, fine-fusion network, and noise reconstruction. Four parts work together to estimate the noise image $I^*_N$. The noise-free data is generated by subtracting $I^*_N$ from the observation data $Y$. The details of each part of the network are presented in the following.

\subsection{Coarse-Fusion Network}

\begin{figure}[htb]
\centering
\includegraphics [width=3.4in]{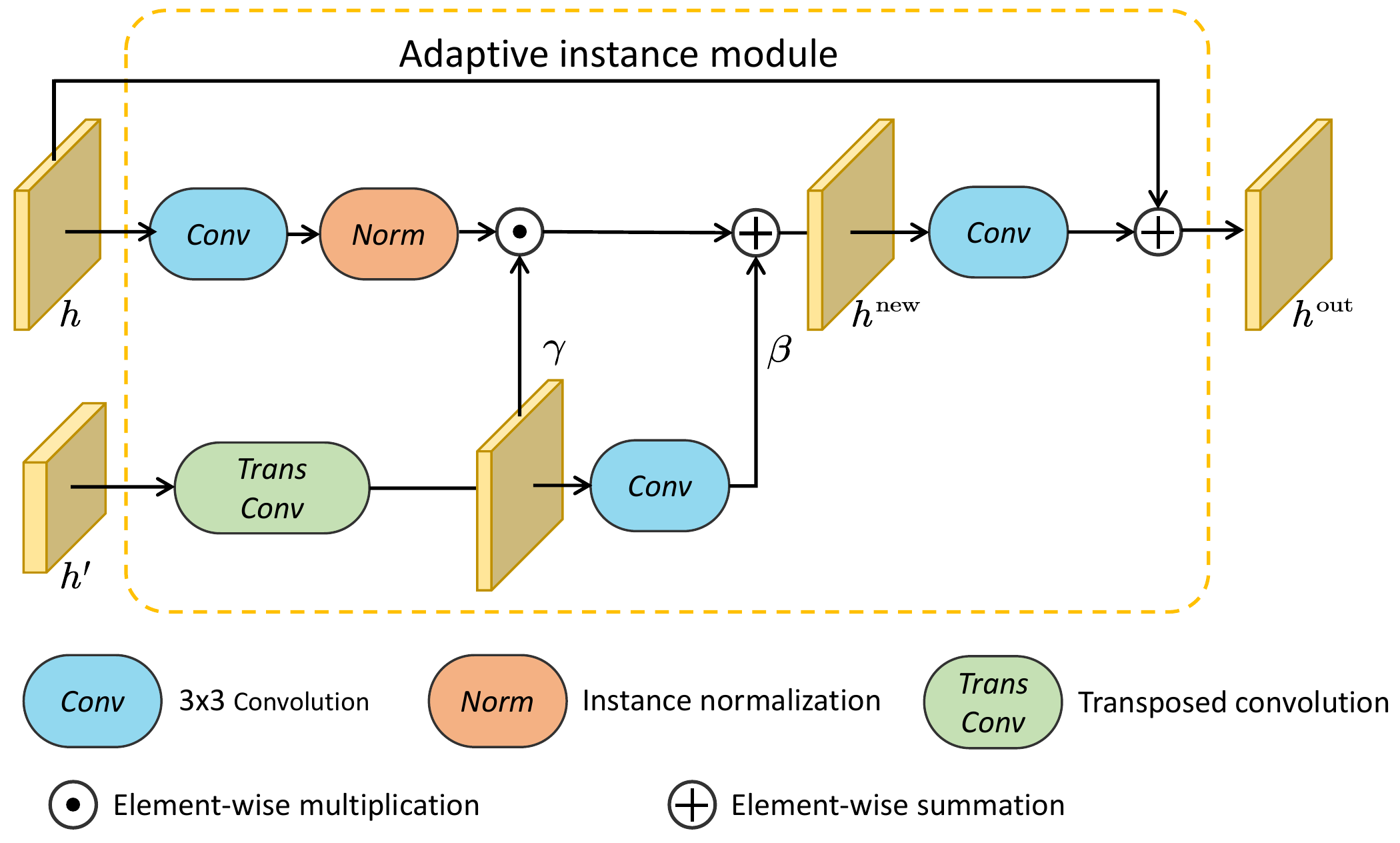}
\caption{Illustration of the proposed adaptive instance (AIN) module.}
\label{fig_ain}
\end{figure}

For a given input hyperspectral image, the proposed model first downsamples the input image into 1/2 and 1/4 scales by using Gaussian kernels. Shallow features are extracted by multiple parallel convolutions, as illustrated in Fig. \ref{fig_framework} (the initial layer). Next, the coarse-fusion network extracts deep features and fuses the multiscale information through several parallel adaptive instance (AIN) modules. The motivations for designing the coarse-fusion network are twofold: 1) The multi-scale structure presents a solution to increase the receptive field to capture more content. 2) The AIN module can transfer the basic structures from the low-resolution feature maps to the high-resolution ones. We choose adaptive instance normalization \cite{huang17iccv} to build the AIN module due to its efficiency and compact representation. Fig. \ref{fig_ain} shows the architecture of the AIN module.

The adaptive instance normalization affine transforms the normalized feature map $h\in \mathbb{R}^{H\times W\times C}$ by taking an input $h'\in \mathbb{R}^{\frac{H}{2} \times \frac{W}{2}\times 2C}$. Here $H$ and $W$ denote the height and width of the feature map, respectively. $C$ is the number of channels. It should be noted that $h'$ is the feature from the downscale. Specifically, the adaptive instance normalization takes the current feature $h$ and the downscale feature $h'$ as input. First, we convert $h'$ to the size of $H\times W\times C$ by transposed convolution to be consistent with the same dimension as $h$. Afterwards, for the purpose of using the contextual semantic information contained in the downscale features, we obtain the affine transform parameters from the transformed $h'$ for each pixel (shift $\beta$ and scale $\gamma$). Every feature map is pixel-wise affine transformed and channel-wise normalized, as illustrated in Fig. \ref{fig_ain}. The updated value in the feature map at position $(i,j,c)$ can be formally represented as:

\begin{equation}
    h^\textrm{new}_{i,j,c}=\gamma_{i,j,c}
    \left( \frac{h_{i,j,c}-\mu_c}{\sigma_c}
    \right) + \beta_{i,j,c},
\end{equation}
where $\mu_c$ denotes the mean of $h$ in channel $c$, $\sigma_c$ denotes the standard deviation of features in channel $c$. To be more specific, $\mu_c$ is computed as:
\begin{equation}
    \mu_c=\frac{1}{HW}\sum^H_i\sum^W_j h_{i,j,c}.
\end{equation}
$\sigma_c$ is computed as:
\begin{equation}
    \sigma^2_c=\frac{1}{HW}\sum^{H}_i\sum^{W}_j(h_{i,j,c}-\mu_c)^2.
\end{equation}

It should be noted that $\gamma_{i,j,c}$ and $\beta_{i,j,c}$ are generated pixel-wisely from $h'$. Therefore, the images with spatially variant noise can be handled adaptively. Finally, a convolutional layer is applied on $h^\textrm{new}$, and residual connection is used to better transfer feature information.

\subsection{Fine-Fusion Network}

The outputs of the coarse-fusion network are fed into the fine-fusion network to refine the information from multiple scales. It is well known in cognitive science that in the primate visual cortex, the local receptive fields of neurons are of different sizes. Hence, the capability of collecting multiscale information should be taken into account in deep networks. 

Inspired by HRNet \cite{wang2020deep}, we conduct repeated multiscale fusion by exchanging information across parallel multiresolution subnetworks. Furthermore, we design a co-attention fusion module to adaptively emphasize informative features from different scales, and therefore enhance the discriminative learning capability of the network for image denoising. 

\begin{figure}[htb]
\centering
\includegraphics [width=3.5in]{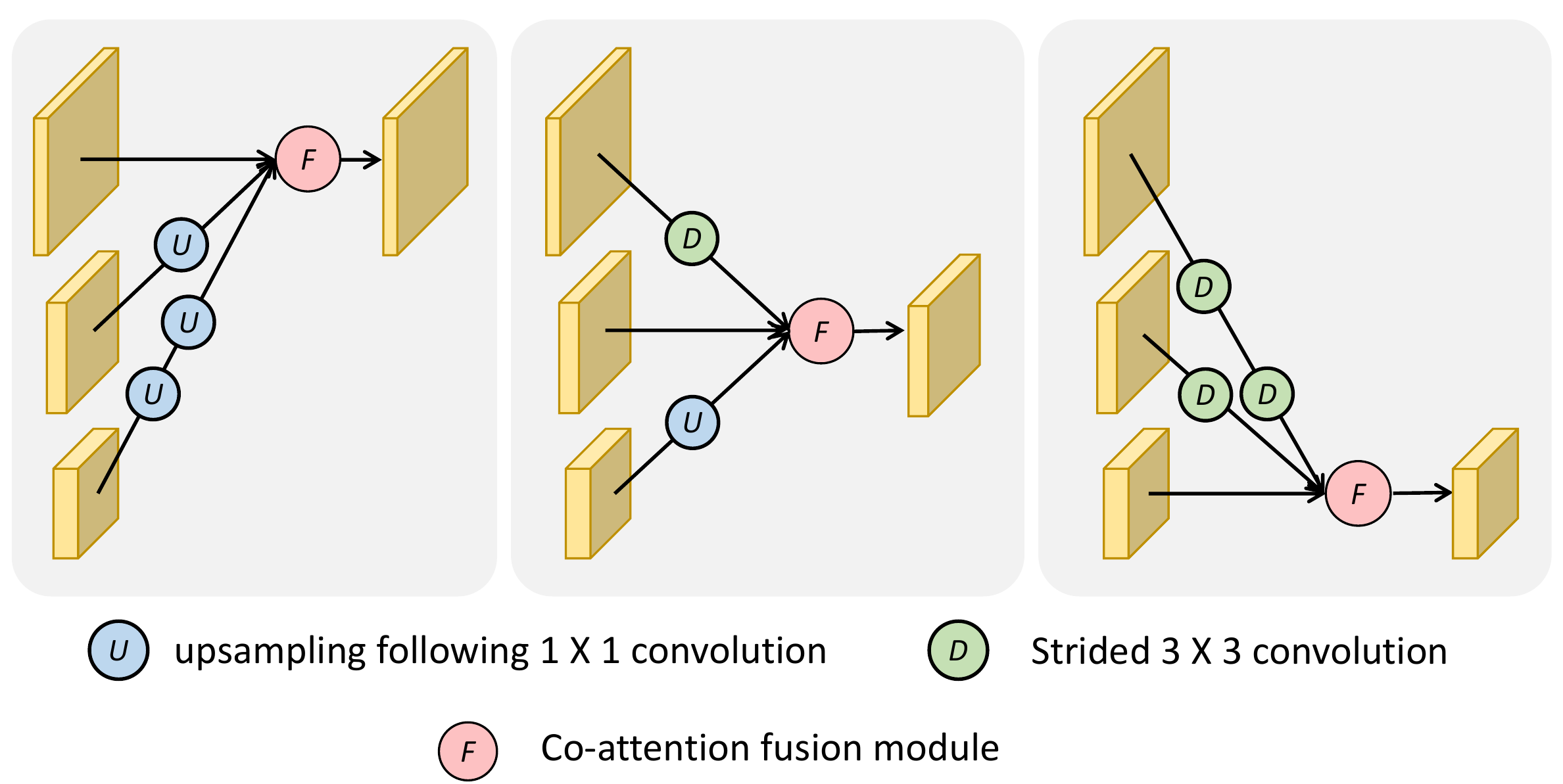}
\caption{Illustration of multiscale features aggregation from left to right, respectively.}
\label{fig_feat_agg}
\end{figure}

As illustrated in Fig. \ref{fig_framework}, the fine-fusion network starts from multiscale feature representations $\{\textbf{X}^1_r, r = 1, 2, 3\}$, where $r$ denotes the spatial resolution index. In the second layer, the feature representations are $\{\textbf{X}^2_r, r = 1, 2, 3\}$. Each feature representation is computed as:
\begin{equation}
    \textbf{X}^2_r=\textrm{CA}(f(\textbf{X}^1_1), f(\textbf{X}^1_2), f(\textbf{X}^1_3)),
\end{equation}
where CA is the co-attention fusion module. $f(\cdot)$ is a transform function. The transform function $f(\cdot)$ depends on the input and output spatial resolution of the feature map. As depicted in Fig. \ref{fig_feat_agg}, the strided $3\times3$ convolution is employed for $2\times$ downsampling. Two consecutive strided $3\times3$ convolutions are employed for $4\times$ downsampling. At the same time, the nearest neighbor sampling following a $1\times1$ convolution is used for upsampling. 

If the input and the output have the same resolution, we adopt the identity connection. Note that after transformation, the feature maps from different scales are of the same size, and they are fed into the co-attention fusion module.

\begin{figure}[htb]
\centering
\includegraphics [width=3.3in]{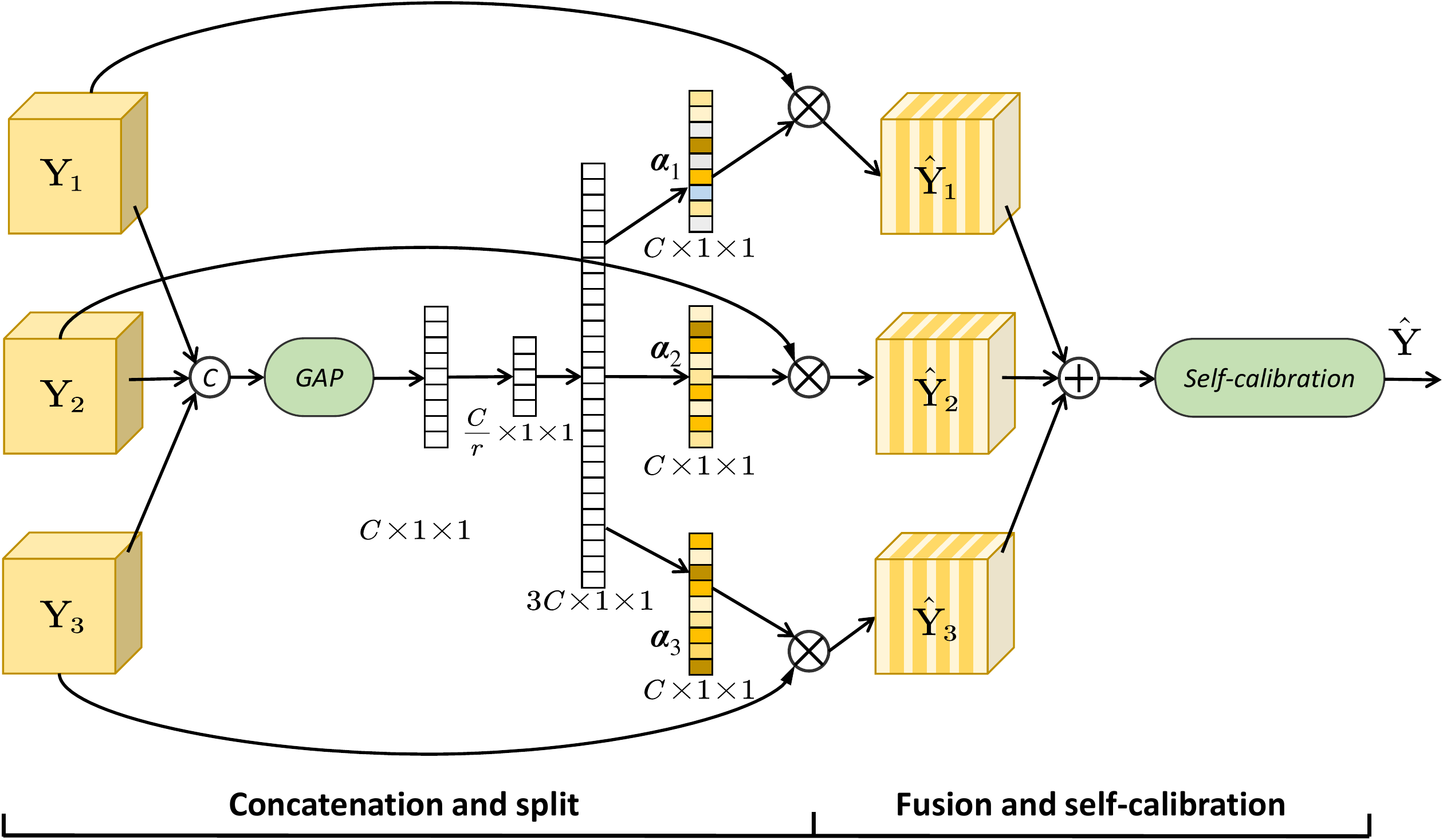}
\caption{Illustration of the proposed co-attention fusion module.}
\label{fig_co_attention}
\end{figure}

In deep neural networks, features from different states or sources contribute differently to the feature representations \cite{9888139}. In HRNet, multiscale features are directly fused by element-wise summation. We argue that simply combining multiscale features by concatenation or summation lacks the flexibility to modulate these features, and the discriminative ability of deep models will be influenced. Therefore, this paper proposes a co-attention fusion module to adaptively emphasize the important information from different scales. The structure of the co-attention fusion module is sketched in Fig. \ref{fig_co_attention}, which consists of two parts: 1) Concatenation and split, 2) Fusion and self-calibration. 

\textbf{Concatenation and split.} The co-attention fusion module receives multiscale features and generates trainable weights for feature fusion. Given input features $\textbf{Y}_1$, $\textbf{Y}_2$, and $\textbf{Y}_3$ are with the size of $C \times H \times W$, we first conduct the concatenation operation on three features:
\begin{equation}
    \textbf{U}=\textrm{cat}(\textbf{Y}_1, \textbf{Y}_2, \textbf{Y}_3),
\end{equation}
where $\textrm{cat}(\cdot)$ is the concatenation operation. Next, The global average pooling is used to compute the channel-wise statistics $\textbf{s}\in \mathbb{R}^{3C\times1\times1}$ along the spatial dimension of $\textbf{U} \in \mathbb{R}^{3C \times H \times W}$. A downsampling convolution layer is employed to produce a compact feature $\textbf{u}\in \mathbb{R}^{\frac{3C}{r}\times1\times1}$. Here $r=4$ is used in our experiments. 

Ultimately, the feature $\textbf{u}$ is passed through three parallel upsampling layers and provides us with three feature descriptors $u_1$, $u_2$ and $u_3$ each with dimension $C\times1\times1$. Softmax function is applied to $u_1$, $u_2$ and $u_3$, yielding three attention activation vectors $\alpha_1$, $\alpha_2$ and $\alpha_3$, respectively.

\textbf{Fusion and self-calibration.} The three attention activation vectors generated will be used to recalibrate the input features as:
\begin{equation}
    \tilde{\textbf{Y}} = \alpha_1 \cdot \textbf{Y}_1 + \alpha_2 \cdot \textbf{Y}_2 + \alpha_3 \cdot \textbf{Y}_3
\end{equation}

Then, we adjust and integrate the features, which are performed by the self-calibration module:
\begin{equation}
    \hat{\textbf{Y}} = H_{sc}(\tilde{\textbf{Y}}),
\end{equation}
where $H_{sc}(\cdot)$ denotes the self-calibrated convolution \cite{liu20cvpr_scconv}. 

As a follow-up operation after fusion, self-calibration convolution uses the convolution filters to operate on the fused feature map to enhance the feature representation ability.

Conclusively, the proposed co-attention module transforms the input features into compact descriptors and generates three sets of weights to model channel-wise interdependencies. In this way, the co-attention module can adaptively emphasize the important information from multiscale and generate trainable weights for representative feature fusion.

\subsection{Denoising and Reconstruction}

At the end of the fine-fusion network, multiscale features are fused by the co-attention fusion module. Then, one convolution layer is employed to learn the residual noise image $I^*_N$. Finally, the noise-free image $\hat{I}$ is computed by subtracting $I^*_N$ from the observation $I_N$. 

We use $L_1$ loss to optimize our network, and the reconstruction loss is:
\begin{equation}
    \mathcal{L}_\textrm{rec} = \Vert \hat{I} - I \Vert_1,
\end{equation}
where $\hat{I}$ denotes the estimated noise-free HSI, and $I$ denotes the real noise-free HSI. While in the HSI case, hundreds of bands with abundant spectral information means the noise types and intensity in each band are usually different. Therefore, differences in spatial and spectral direction can provide additional complementary contributions for denoising. We introduce a global gradient regularizer to constrain the details of $\hat{I}$,
\begin{equation}
\begin{split}
	\mathcal{L}_{grad} = 
	\Vert \nabla_h\hat{I} - \nabla_{h}I \Vert _2^2 + \Vert \nabla_v\hat{I} - \nabla_{h}I \Vert _2^2 + \\
	\Vert \nabla_s\hat{I} - \nabla_{s}I \Vert _2^2 ~~~~~~~~~~~~~~~~~~~~~~~~
\end{split}
\end{equation}
where $\nabla_h$, $\nabla_v$ and $\nabla_s$ denote the gradient operator along the horizontal, vertical, and spectral direction respectively. Then, the total loss function is as follows:
\begin{equation}
	\mathcal{L} = \mathcal{L}_{rec} + \lambda \mathcal{L}_{grad},
\end{equation}
where $\lambda$ is the weight parameter of $\mathcal{L}_{grad}$. We empirically set $\lambda$ to 0.01 to balance the loss terms.

\section{Experimental Results and Analysis}

\subsection{Experiment Setup}

\textbf{Benchmark datasets.} To verify the effectiveness of the proposed MAFNet for HSI denoising. The proposed MAFNet is employed on several datasets, and training is conducted using data from ICVL \cite{arad16eccv} and CAVE \cite{CAVE_0293} hyperspectral dataset. The images in the ICVL dataset were collected over 31 spectral bands with the size of $1392\times 1300$, while the images in CAVE dataset were collected over 31 spectral bands with the size of $512\times512$. The training data are randomly cropped as cube data with the size of $128\times128\times31$. Basic data augmentation (rotation and scaling) is used for regularization. Twenty thousand training samples are generated in total. To verify the robustness of the proposed MAFNet in real data, spaceborne hyperspectral data are used in our experiments, including Pavia University, Urban and Indian Pines. Through experiments on both real noise HSI datasets, we try to verify the generalization ability and denoising effect of the proposed MAFNet.

\textbf{Noise setting.} Hyperspectral data captured by real spaceborne sensors are commonly contaminated by a mixture noise, such as the Gaussian noise, impulse noise, and deadline noise. In the testing phase, five types of complex noise are defined as follows:

\begin{enumerate}[]

\item \textit{Case 1: Non-i.i.d. Gaussian noise.} Data in all spectral bands are contaminated by Gaussian noise with various intensities. The variances of Gaussian noise are randomly selected from 30 to 70.

\item \textit{Case 2: Gaussian + Stripe noise.} Every band is contaminated by \textit{non-i.i.d Gaussian noise}, as mentioned in Case 1. Besides, some spectral bands are randomly selected to add strip noise. In each band, 5\% to 15\% of columns are polluted with strips.

\item \textit{Case 3: Gaussian + Deadline noise.} Every band is corrupted by \textit{non-i.i.d Gaussian noise}, as mentioned in Case 1. Besides, deadline noise is randomly added to one-third of spectral bands. In each band, 5\% to 15\% of columns are conflicted with deadlines.

\item \textit{Case 4: Gaussian + Impulse noise.} Each band is contaminated by Gaussian noise, as mentioned in Case 1. One-third of bands are randomly selected to add impulse noise with intensity ranging from 10\% to 70\%.

\item \textit{Case 5: Mixture noise.} Like other cases, every spectral band is corrupted by Gaussian noise. Then, each band is randomly contaminated by a random combination of the other three noises.

\end{enumerate}

\textbf{Competing methods and quantitative metrics.} The proposed method was compared with six state-of-the-art methods. Both traditional methods and deep learning-based methods are taken into account. Specifically, For traditional methods, BM4D \cite{maggioni13tip}, low-rank methods (LRMR \cite{zhang14tgrs} and LRTV \cite{he16tgrs}) are considered. For deep learning-based methods, the proposed MAFNet is compared with HSID-CNN \cite{yuan19tgrs}, MemNet \cite{tai17iccv} and QRNN3D \cite{wei21tnnls}. To give a fair evaluation, three quantitative metrics are used, including peak signal-to-noise ratio (PSNR), structure similarity (SSIM) \cite{wang04tip}, and spectral angle mapper (SAM) \cite{yuhas93jpl}. SAM is a spectral-based index, and a smaller value of SAM indicates better denoising performance. PSNR and SSIM are spatial-based indexes. Larger values of PSNR and SSIM suggest better denoising performance.

\begin{figure}[htb]
\centering
\includegraphics[width=0.45\textwidth]{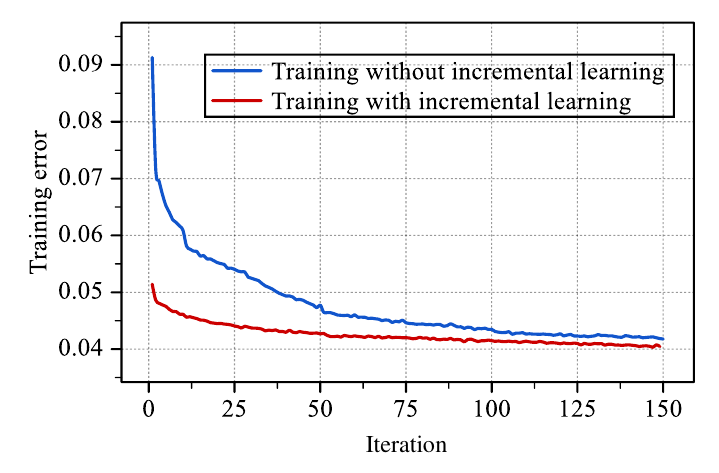}
	\caption{The training errors with / without incremental learning. Blue curve denotes the mixture noise data training without incremental learning. Red curve denotes the mixture noise data training with incremental learning.}
	\label{learning policy}
\end{figure}

\textbf{Incremental learning policy.} We use an incremental learning policy for stable training, which can effectively avoid the network converging to suboptimal minimum. Specifically, the training goes through three stages, and the training data of each phase uses the same network. In the first stage, Gaussian noise with fixed noise level ($\sigma =$ 30, 50, 70) is sequentially employed to build the training data for network training in turn. The network weights of each training phase are saved, and we load the last trained network weights to initialize the network parameters for the next training phase, instead of retraining the network again from scratch. Next, we use blind Gaussian noise (randomly selected from $\sigma = 30, 50, 70$) to construct the training data, and the method described in the first stage is still used to load the network weight data already trained in the first stage. Finally, the complex noise is employed to produce the training data (from Case 1 to Case 5). With the increase of the noise complexity of each stage, the denoising difficulty of the model also increases. Therefore, making full use of the pre-trained network model in the previous stage is more conducive to improve denoising performance and enables the network to converge better. In order to explore the effectiveness of the incremental learning strategy, we plotted the loss function curve in the model training process, as shown in Fig. \ref{learning policy}, we can see that the incremental learning strategy can converge faster and achieve better denoising performance.

It should be noted that we make the model handle data with different noises sequentially. Following the easy-to-difficult learning strategy in Curriculum learning \cite{bengio2009curriculum}, we incrementally learning the noise from Case 1 to Case 4, and therefore gradually improving the generalization and learning ability of the model. Finally, the model learns the complex noise, including all kinds of noise from Case 1 to Case 4. Hence, the final model is robust for complex noise removal. Through incremental learning, the proposed MAFNet achieves better denoising performance.

We initialize the learning rate at $10^{-4}$, and it decayed every epoch to accelerate training. The training process of MAFNet took 100 epochs for Gaussian noise and 150 epochs for complex noise. The network is optimized using the Adam  optimizer with the PyTorch framework on a machine with NVIDIA GTX 2080Ti GPU, Intel(R) Xeon(R) E5 CPU of 2.50GHz and 32GB RAM.

\begin{figure*}[htb]
	\centering
	\includegraphics[width=0.8\textwidth]{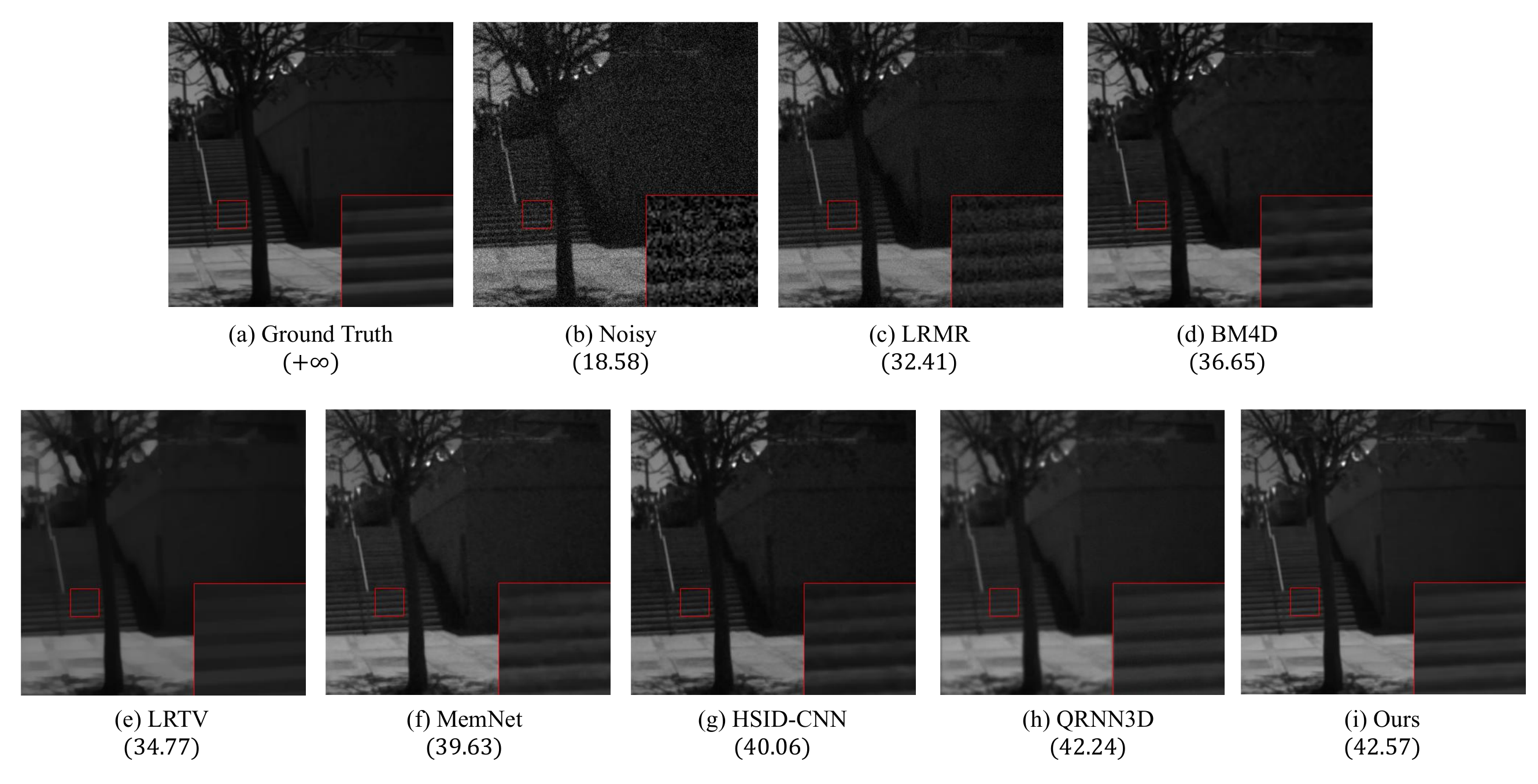}
	\caption{The Gaussian noise removing results of PSNR (dB) at $10^{th}$ band of the HSI under noise level $\sigma=30$ (ICVL dataset).}
	\label{gauss30}
\end{figure*}

\renewcommand\arraystretch{1.4}
\begin{table*}[htb]
\centering
\caption{Quantitative results of different methods on ICVL dataset. ``Blind" denotes that each image is corrupted by Gaussian noise with unknown $\sigma$.}
\begin{tabular}{cc|cccccccc} \hline
\toprule
	~~~Case~~~ & ~~~Index~~~ & ~Noisy~  & ~LRMR \cite{zhang14tgrs}~ & ~BM4D \cite{maggioni13tip}~ & ~LRTV \cite{he16tgrs}~ & ~MemNet \cite{tai17iccv}~ & HSID-CNN \cite{yuan19tgrs} & QRNN3D \cite{wei21tnnls} & ~Ours~~~\\
\midrule
	\multirow{3}*{$\sigma = 30$} 
	& PSNR $\uparrow$ & 18.58 & 31.73 & 36.72 & 34.79 & 40.17 & 41.14 & 43.46 & \textbf{43.97} \\
	~ & SSIM $\uparrow$ & 0.107 & 0.679 & 0.865 & 0.769 & 0.962 & 0.957 & 0.973 & \textbf{0.975} \\
	~ & SAM $\downarrow$ & 0.703 & 0.179 & 0.143 & 0.157 & 0.074 & 0.096 & 0.038 & \textbf{0.026} \\
\midrule
	\multirow{3}*{$\sigma = 50$} 
	& PSNR $\uparrow$ & 14.15 & 29.56 & 34.92 & 32.75 & 37.68 & 37.21 & 41.93 & \textbf{42.03} \\
	~ & SSIM $\uparrow$ & 0.043 & 0.634 & 0.769 & 0.646 & 0.941 & 0.932 & 0.957 & \textbf{0.963} \\
	~ & SAM $\downarrow$ & 0.890 & 0.214 & 0.192 & 0.194 & 0.093 & 0.124 & 0.040 & \textbf{0.029} \\
\midrule
	\multirow{3}*{$\sigma = 70$} 
	& PSNR $\uparrow$ & 11.22 & 26.44 & 31.78 & 29.13 & 35.83 & 35.46 & 39.73 & \textbf{40.84} \\
	~ & SSIM $\uparrow$ & 0.023 & 0.576 & 0.601 & 0.598 & 0.897 & 0.896 & 0.945 & \textbf{0.954} \\
	~ & SAM $\downarrow$ & 1.013 & 0.282 & 0.231 & 0.252 & 0.097 & 0.148 & 0.051 & \textbf{0.031} \\
\midrule
	\multirow{3}*{blind}
    & PSNR $\uparrow$ & 14.66 & 30.16 & 33.90 & 32.26 & 38.84 & 38.32 & 42.03 & \textbf{42.15} \\
	~ & SSIM $\uparrow$ & 0.058 & 0.652 & 0.835 & 0.774 & 0.953 & 0.943 & 0.960 & \textbf{0.963} \\
	~ & SAM $\downarrow$ & 0.868 & 0.196 & 0.121 & 0.151 & 0.082 & 0.131 & 0.044 & \textbf{0.029} \\

\bottomrule
\end{tabular}
	\label{gauss case}
\end{table*}

\begin{figure*}[htb]
	\centering
	\includegraphics[width=0.8\textwidth]{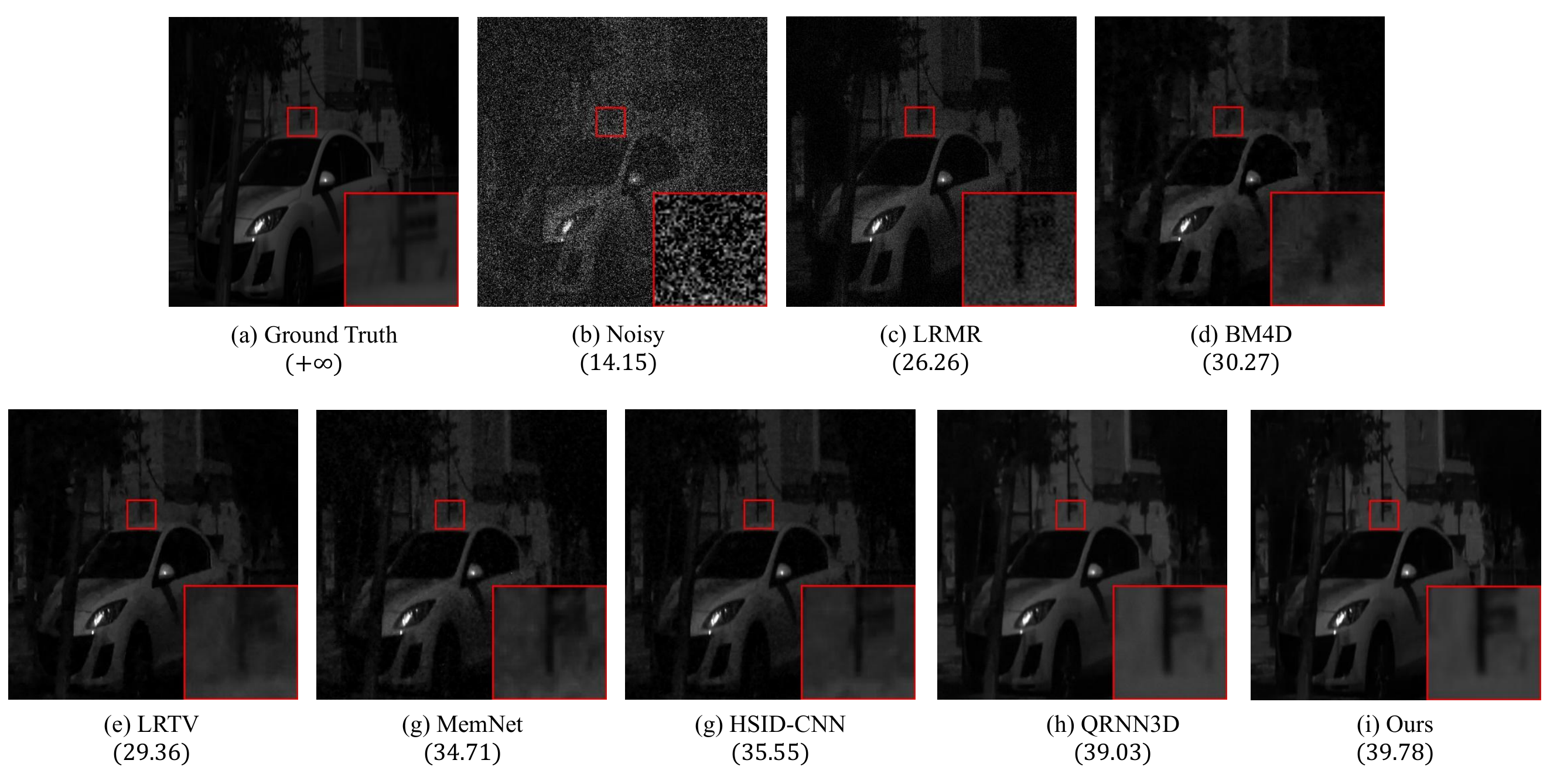}
	\caption{Gaussian noise removal results of PSNR (dB) at $10^{th}$ band of image under noise level $\sigma=70$ (ICVL dataset).}
\label{gauss70}
\end{figure*}

\subsection{Experiments on Gaussian Noise Cases}

This paper uses a single model to process various Gaussian noise levels. Specifically, additive Gaussian noise with different variances is imposed to produce a set of noisy HSI patches. The average evaluation indexes are listed in Table \ref{gauss case}. The best performance for each quality index is marked in bold. Fig. \ref{gauss30} and \ref{gauss70} show the denoising results under noise levels $\sigma = 30$ and $\sigma = 70$ to give detailed comparison results.

Through comparison, we can observe that the proposed MAFNet obtains better performance metrics (PSNR, SSIM, and SAM) when dealing with Gaussian noise cases. It is owing to this reason that the proposed MAFNet takes the multi-scale contextual information into account. Furthermore, benefiting from the AIN module and co-attention fusion, the MAFNet also achieves better denoising results compared with HSID-CNN, MemNet and QRNN3D. As shown in Fig. \ref{gauss30} and \ref{gauss70}, we select one band to give the denoising results. It can be easily seen that the denoising result of the proposed MAFNet is capable of effectively reducing the Gaussian noise while precisely preserving the basic texture details of the original HSI. LRMR can hardly deal with degraded bands with strong Gaussian noise. BM4D introduces undesirable artifacts in some regions. Other deep learning-based methods perform better in noise suppression, but still lose some texture details. Furthermore, the PSNR and SSIM values of other deep learning-based methods are lower than the proposed MAFNet.

\begin{figure*}
\centering
\includegraphics[width=6.8in]{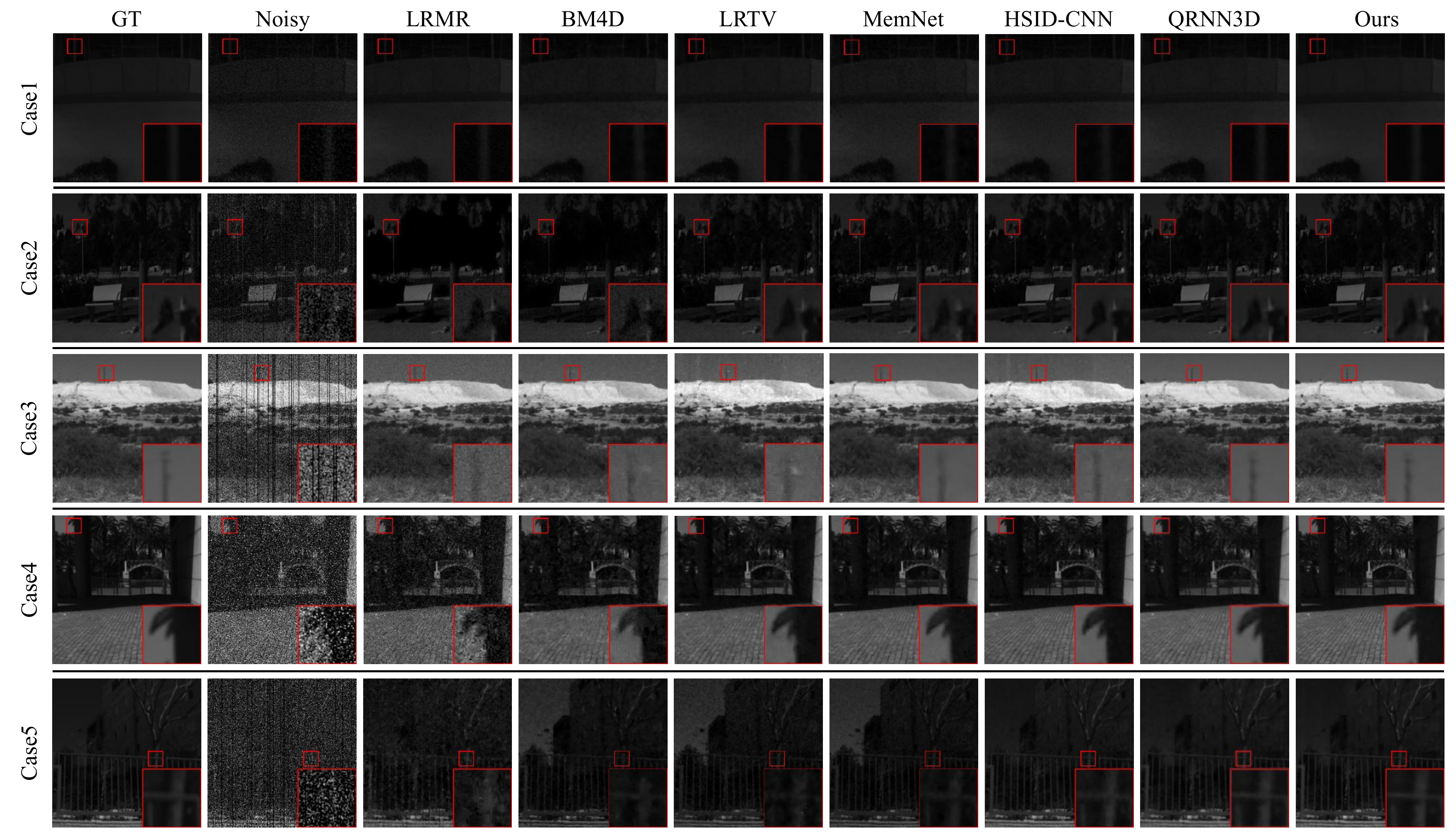}
\caption{Complex noise removal results on the ICVL dataset. Examples for non-i.i.d Gaussian noise (Case 1), Gaussian + stripes (Case 2), Gaussian + deadline (Case 3), Gaussian + impulse (Case 4) and mixture noise (Case 5) removal are illustrated, respectively.}
\label{complex fig}
\end{figure*}

\begin{table*}[htb]
\renewcommand\arraystretch{1.4}
\centering
\caption{Quantitative results of different methods under five complex noise cases on ICVL dataset.}
\begin{tabular}{cc|cccccccc}
\toprule
	~~~Case~~~ & ~~~Index~~~ & ~Noisy~  & ~LRMR \cite{zhang14tgrs}~ & ~BM4D \cite{maggioni13tip}~ & ~LRTV \cite{he16tgrs}~ & ~MemNet \cite{tai17iccv}~ & HSID-CNN \cite{yuan19tgrs} & QRNN3D \cite{wei21tnnls} & ~Ours~~~\\
\midrule
	\multirow{3}*{Case 1} 
	& PSNR $\uparrow$ & 17.99 & 32.20 & 35.46 & 36.12 & 41.32 & 41.14 & 42.51 & \textbf{43.54} \\
	~ & SSIM $\uparrow$ & 0.160 & 0.692 & 0.892 & 0.905 & 0.968 & 0.952 & 0.969 & \textbf{0.974} \\
	~ & SAM $\downarrow$  & 0.853 & 0.201 & 0.137 & 0.115 & 0.050 & 0.057 & 0.043 & \textbf{0.028} \\ 
\midrule
	\multirow{3}*{Case 2} 
	& PSNR $\uparrow$ & 17.62 & 30.87 & 34.17 & 34.89 & 40.32 & 39.68 & 41.78 & \textbf{42.50} \\
	~ & SSIM $\uparrow$ & 0.151 & 0.653 & 0.821 & 0.890 & 0.954 & 0.939 & 0.960 & \textbf{0.968} \\
	~ & SAM $\downarrow$  & 0.855 & 0.257 & 0.159 & 0.129 & 0.078 & 0.071 & 0.049 & \textbf{0.030} \\
\midrule
	\multirow{3}*{Case 3} 
	& PSNR $\uparrow$ & 17.53 & 30.17 & 33.65 & 34.02 & 39.26 & 39.75 & 40.88 & \textbf{41.77} \\
	~ & SSIM $\uparrow$ & 0.150 & 0.660 & 0.860 & 0.847 & 0.947 & 0.932 & 0.954 & \textbf{0.966} \\
	~ & SAM $\downarrow$  & 0.871 & 0.249 & 0.174 & 0.160 & 0.081 & 0.091 & 0.059 & \textbf{0.032} \\
\midrule
	\multirow{3}*{Case 4} 
	& PSNR $\uparrow$ & 15.05 & 29.37 & 32.75 & 33.14 & 37.22 & 36.83 & 37.71 & \textbf{37.77} \\
	~ & SSIM $\uparrow$ & 0.117 & 0.644 & 0.801 & 0.795 & 0.911 & 0.898 & 0.938 & \textbf{0.946} \\
	~ & SAM $\downarrow$  & 0.900 & 0.271 & 0.196 & 0.182 & 0.114 & 0.138 & 0.095 & \textbf{0.067} \\
\midrule
	\multirow{3}*{Case 5} 
	& PSNR $\uparrow$ & 14.18 & 27.16 & 30.81 & 31.10 & 36.14 & 36.69 & 37.04 & \textbf{37.19} \\
	~ & SSIM $\uparrow$ & 0.095 & 0.591 & 0.719 & 0.735 & 0.905 & 0.884 & 0.922 & \textbf{0.938} \\
	~ & SAM $\downarrow$  & 0.918 & 0.312 & 0.224 & 0.211 & 0.135 & 0.167 & 0.105 & \textbf{0.065} \\	
\bottomrule
\end{tabular}
\label{complex table}
\end{table*}

\begin{table*}[htb]
\renewcommand{\arraystretch}{1.4}
	\centering
	\caption{Quantitative results of different methods on the Pavia University dataset.}
	\begin{tabular}{cc|ccccccc} \hline
	\toprule
		~~Index~~~ & ~~~Noisy~~~ & ~~~LRMR \cite{zhang14tgrs}~~~ & ~~~BM4D \cite{maggioni13tip}~~~ &  ~~~LRTV \cite{he16tgrs}~~~ & ~~~MemNet \cite{tai17iccv}~~~ & HSID-CNN \cite{yuan19tgrs} & QRNN3D \cite{wei21tnnls} & ~~~Ours~~~~~ \\ 
	\midrule
		 PSNR $\uparrow$ & 17.99  & 27.64 & 25.16  & 26.75 & 29.27 & 30.38 & 33.83 & \textbf{34.03} \\ 
		 SSIM $\uparrow$ & 0.160 & 0.717 & 0.613 & 0.675 & 0.794 & 0.809 & 0.894 & \textbf{0.912} \\ 
		 SAM $\downarrow$  & 0.853 &0.308 & 0.397 & 0.327 & 0.137 & 0.148 & 0.109 & \textbf{0.101} \\ 		
	\bottomrule
	\end{tabular}
	\label{pavia table}
\end{table*}

\subsection{Experiments of Complex Noise Removal on ICVL Dataset}

As mentioned before, the model trained at the final stage of training is used to deal with the five complex noise cases simultaneously. Five types of complex noise include Non-i.i.d Gaussian noise, Gaussian + stripe noise, Gaussian + deadline noise, Gaussian + impulse noise, and Mixture noise. We conducted experiments on ICVL dataset for complex noise removal. The quantitative results are shown in Fig. \ref{complex fig}, and the corresponding quantitative values are listed in Table \ref{complex table}, respectively.

\begin{figure*}
\centering
\includegraphics[width=6.2in]{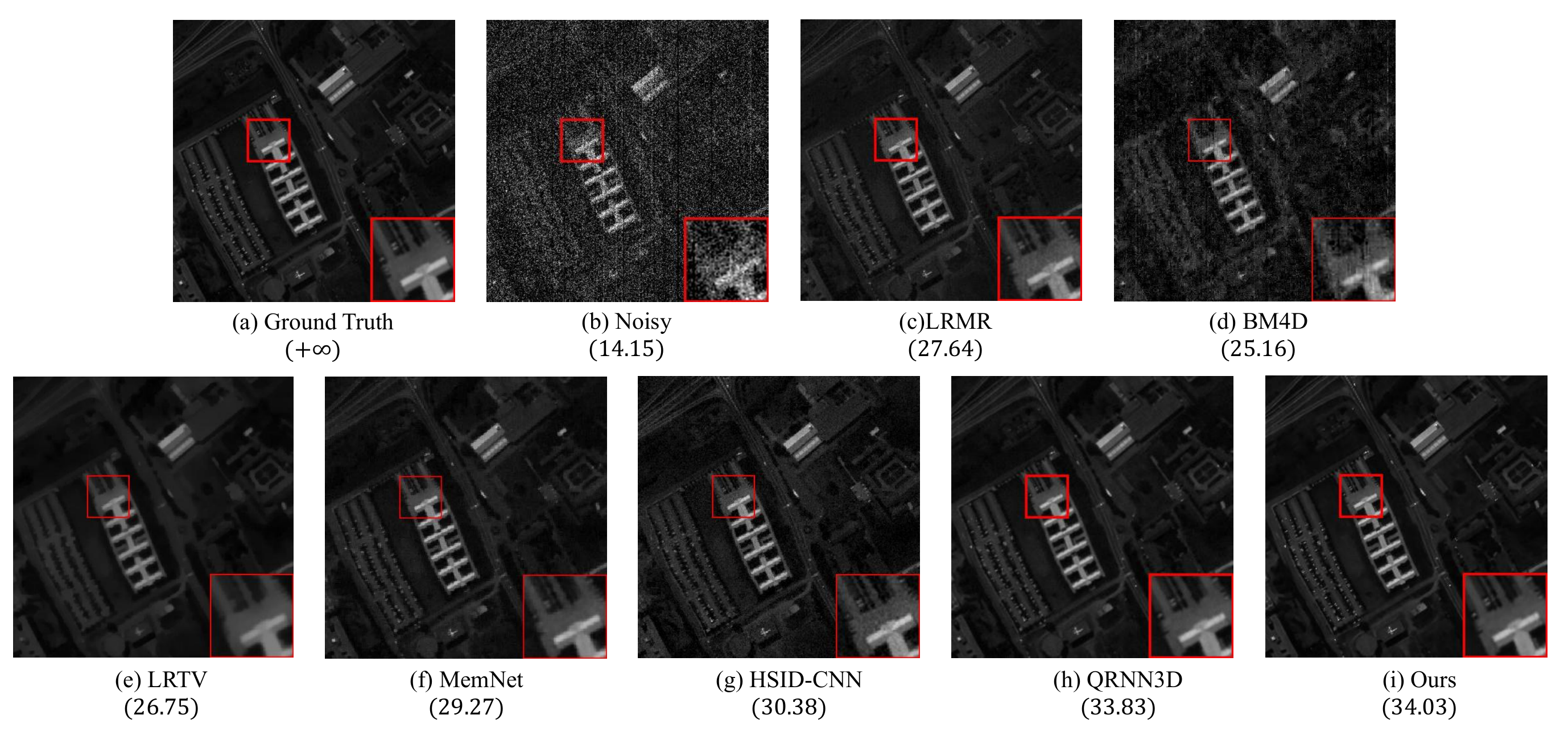}
\caption{Simulated noise removal results of PSNR (dB) at $10^{th}$ band of mixture noise on the Pavia University dataset.}
\label{pavia fig}
\end{figure*}

\begin{figure*}[htbp]
\centering
\subfigure[]{\includegraphics[width=.45\textwidth]{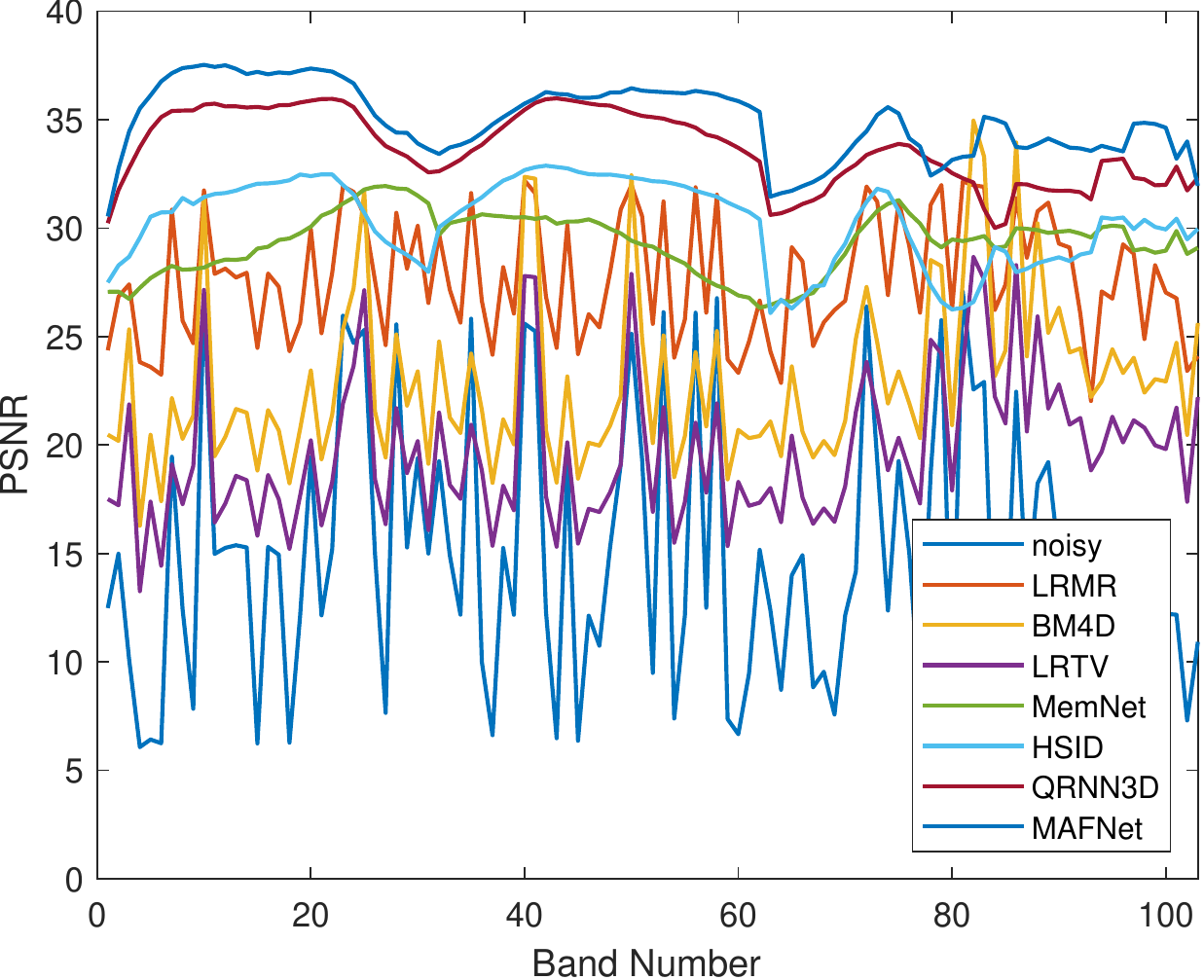}}
\subfigure[]{\includegraphics[width=.45\textwidth]{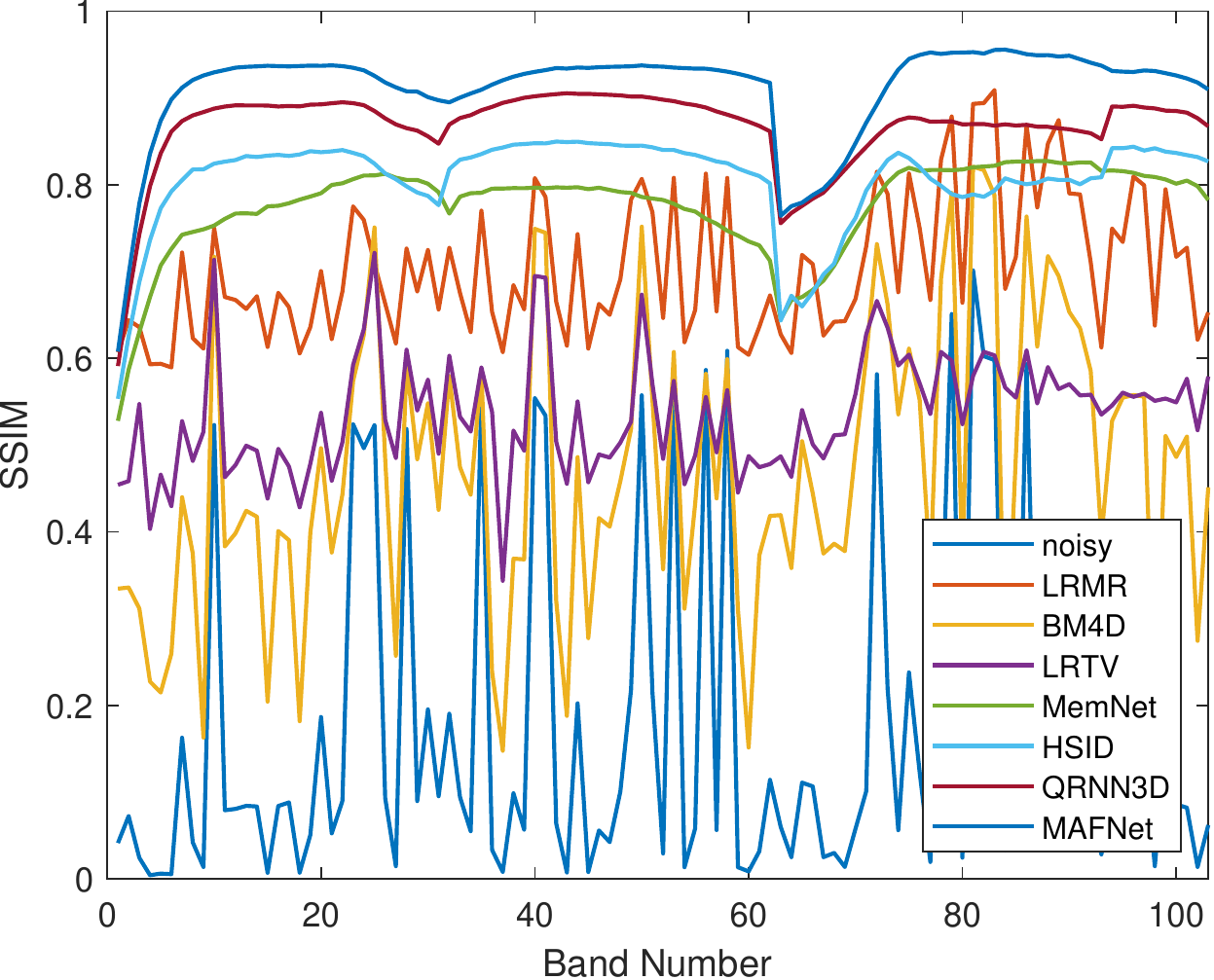}}
\caption{PSNR and SSIM values of different denoising methods in each band of the simulated experiments in mixture noise on Pavia University}
\label{value}
\end{figure*}

\begin{figure*}
\centering
\includegraphics[width=.95\textwidth]{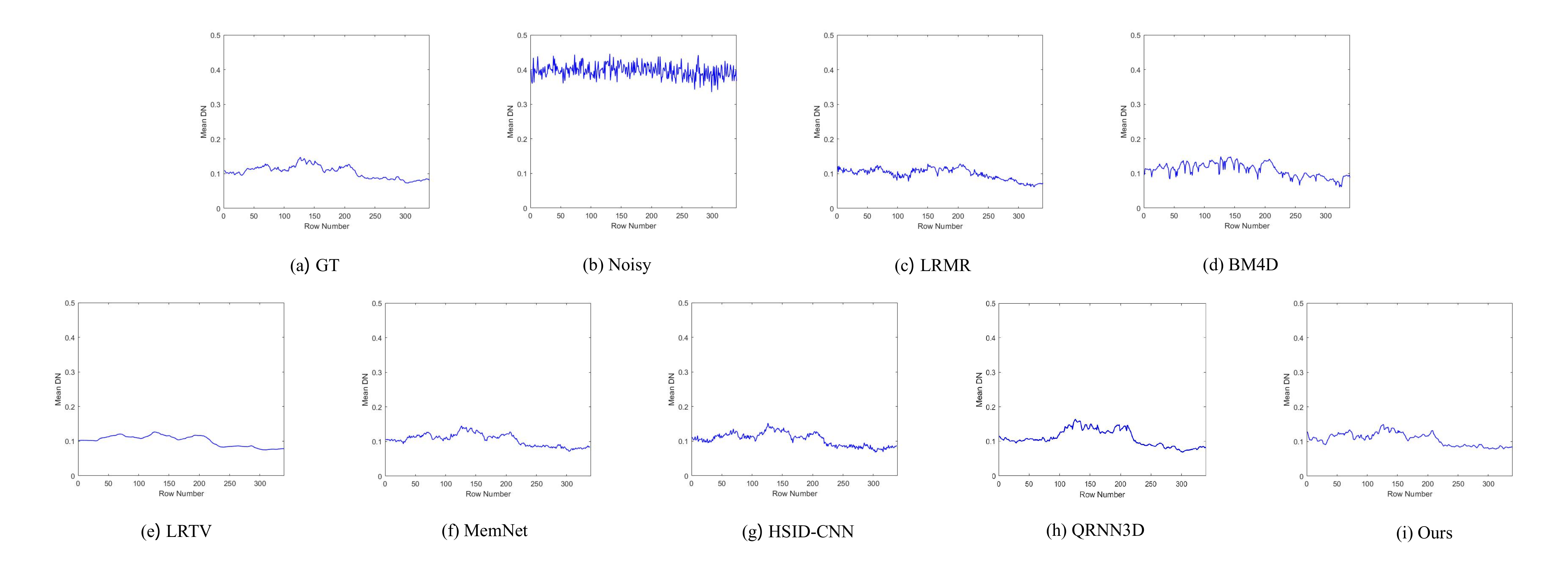}
\caption{Horizontal mean digital number curves at $14^{th}$ band of real noise by different methods on the Pavia University dataset.}
\label{rdn}
\end{figure*}

Fig. \ref{complex fig} shows the visual results of MAFNet denoising under complex noise conditions. The result shows that our MAFNet significantly outperforms the other methods. By comparing the denoising performance indicators in Table \ref{complex table}, it can be observed that our MAFNet performs better than LRMR and LRTV, since they are low-rank matrix-based methods and some basic structures get lost in the denoising process. Furthermore, our MAFNet performs better than the other deep learning-based methods (MemNet, HSID-CNN and QRNN3D). It is evident that the multi-scale feature exploitation and contextual information integration can help the network to capture more intrinsic characteristics of HSI. At the same time, it also helps the network to retain more structural information about the input image. As shown in Fig. \ref{complex fig}, our MAFNet not only removes complex noise, but also retains the structure and spatial details. Moreover, compared with other methods, the proposed MAFnet generates HSI images with a more natural and vivid appearance. Besides, the HSI images produced by the proposed MAFNet have better global contrast.

\subsection{Experiments of Complex Noise Removal on CAVE Dataset}

We conducted complex noise removal experiments on the CAVE dataset \cite{CAVE_0293} so as to verify the effectiveness of MAFNet's denoising performance. Each image in the dataset is acquired at a wavelength of 10 nm in the range of 400 - 700 nm. We divided the dataset into two sets, with 20 images for training and 12 images for testing. Table \ref{cave table} presents the quantitative evaluation results of different methods. We compare the proposed MAFNet with BM4D \cite{maggioni13tip}, OLRT \cite{chang2020hyperspectral}, NGMeet \cite{he2020non}, MemNet \cite{tai17iccv}, HSID-CNN \cite{yuan19tgrs} and QRNN3D \cite{wei21tnnls}. It can be observed that the proposed MAFNet achieves the highest PSNR value, which demonstrates its superior denoising performance. 

It should be noted that we employ state-of-the-art low-rank tensor recovery models (NGMeet \cite{he2020non} and OLRT \cite{chang2020hyperspectral}) on the CAVE dataset. These low-rank tensor methods can effectively utilize both the spatial-spectral information, and preserve the high-dimensional spatial and spectral structure information in HSIs. NGMeet learns the orthogonal basis matrix and reduced image, which produces impressive recovered images. OLRT provides a flexible and promising solution for HSI denoising, deblurring, and inpainting. Additionally, we conducted experiments of Gaussian noise on CAVE dataset. In the low noise case, NGMeet and OLRT demonstrate better performance than MAFNet. As the noise level increases, the proposed MAFNet outperforms NGMeet and OLRT. Therefore, we can conclude that NGMeet and OLRT are effective in spectral regularization, and the proposed MAFNet is beneficial for complex noise removal.

\begin{table*}[htb]
\renewcommand{\arraystretch}{1.4}
\centering
\caption{Quantitative results of different methods on the CAVE dataset.}
\begin{tabular}{cc|ccccccc} \hline
\toprule
~~Index~~~ & ~~~Noisy~~~ & ~~~BM4D \cite{maggioni13tip}~~ & ~~~OLRT \cite{chang2020hyperspectral}~~ &  ~~~NGMeet \cite{he2020non}~~ & MemNet \cite{tai17iccv} & ~~~HSID-CNN \cite{yuan19tgrs}~~ & QRNN3D \cite{wei21tnnls} & ~~~Ours~~~~ \\ 
\midrule
PSNR $\uparrow$ & 12.92  & 29.07 & 31.09 & 32.74 & 32.04 & 32.82 & 33.94 & \textbf{34.19} \\ 
SSIM $\uparrow$ & 0.092 & 0.704 & 0.785 & 0.792 & 0.854 & 0.890 & 0.914 & \textbf{0.931} \\ 
SAM $\downarrow$  & 1.203 & 0.384 & 0.164 & 0.152 & 0.119 & 0.101 & 0.095 & \textbf{0.081} \\ 
\bottomrule
\end{tabular}
\label{cave table}
\end{table*}

\begin{figure*}
\centering
\includegraphics[width=6.8in]{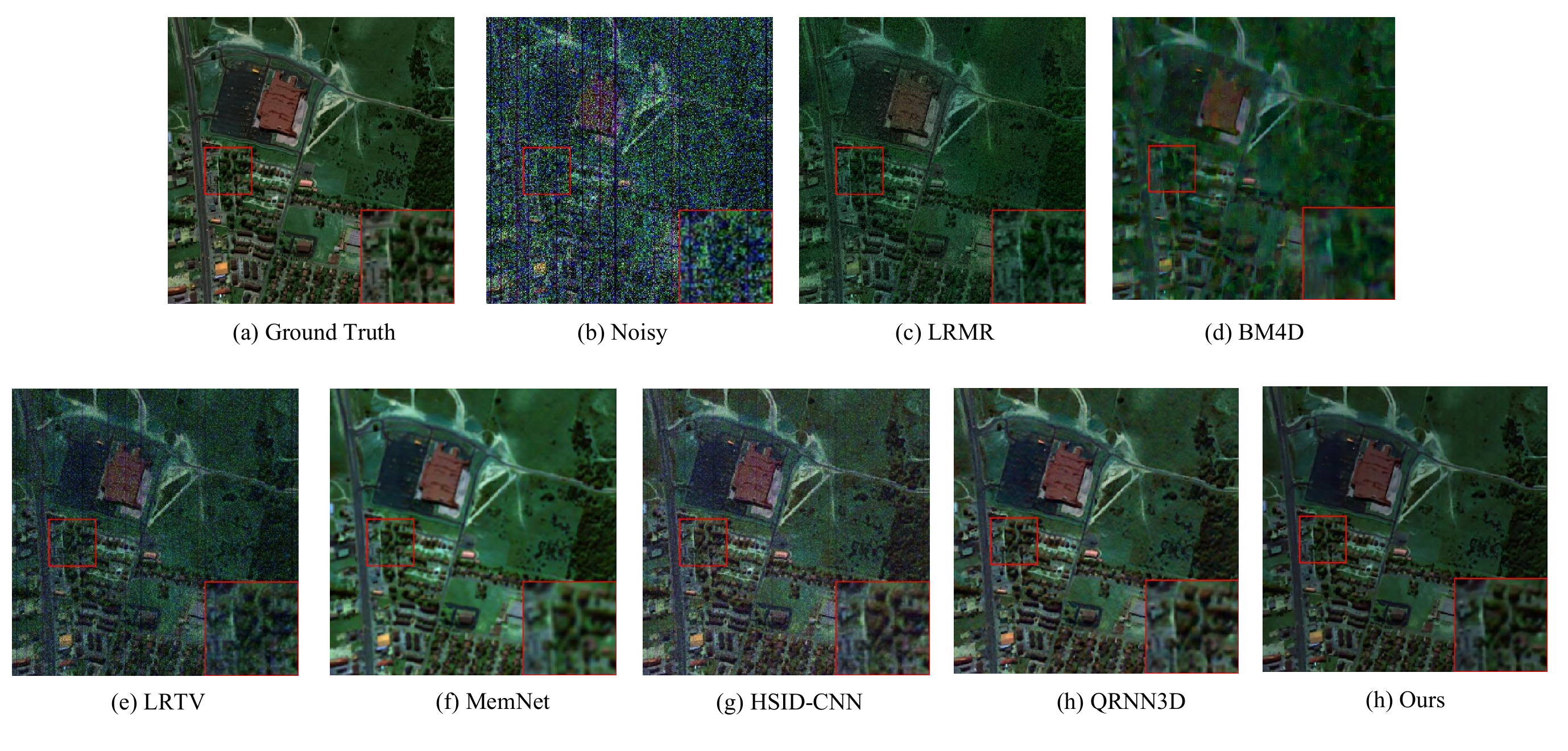}
\caption{Noise removal visual results of false color images on band 50, 100 and 150 of the Urban dataset with competing methods.}
\label{urban fig}
\end{figure*}

\begin{figure*}
\centering
\includegraphics[width=6.8in]{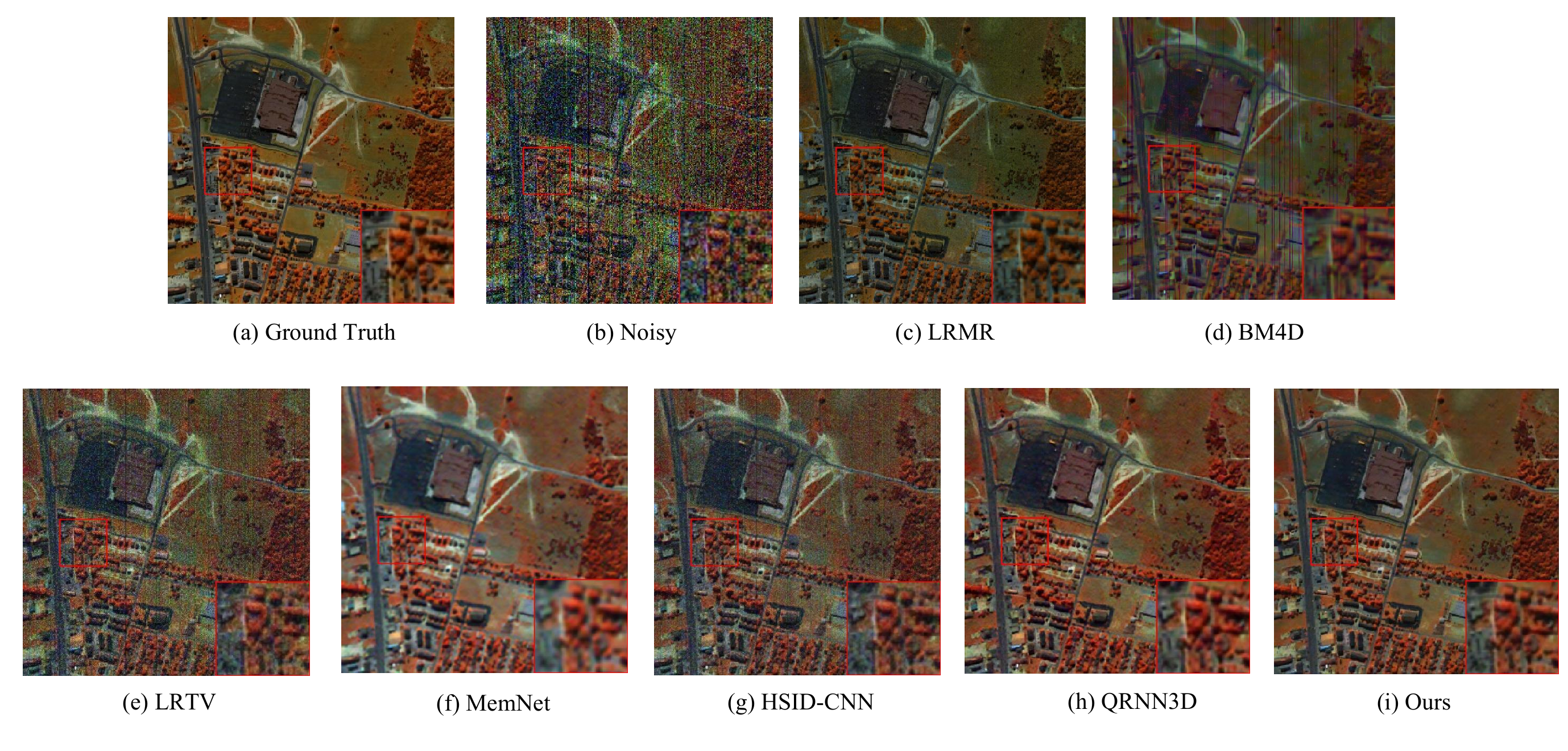}
\caption{Noise removal visual results of false color images on band 70, 110 and 160 of the Urban dataset with competing methods.}
\label{urban_2 fig}
\end{figure*}

\begin{figure*}
\centering
\includegraphics[width=6.8in]{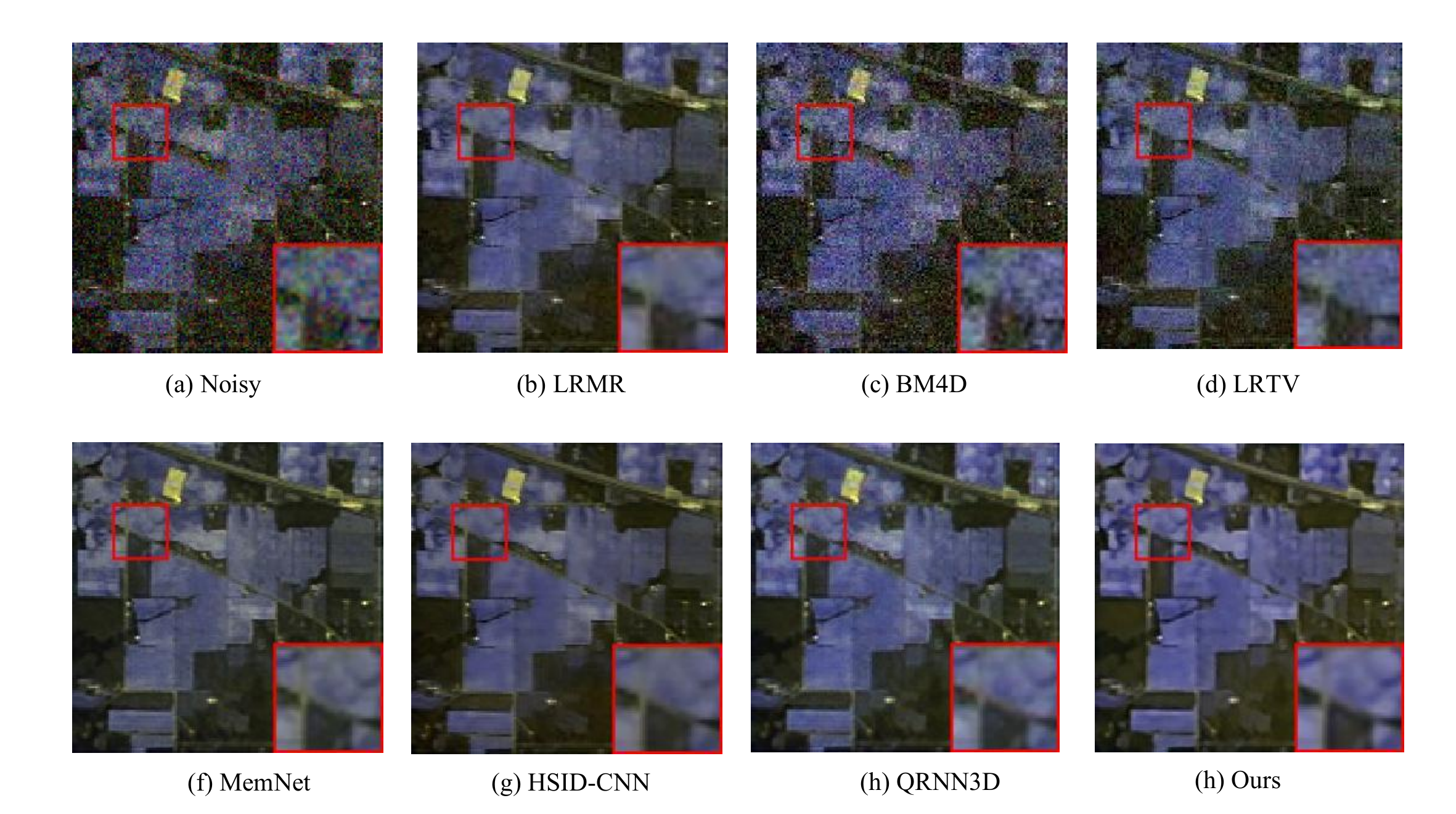}
\caption{Noise removal visual results of false color images on band 2, 3 and 203 of the India Pines dataset with competing methods.}
\label{india fig}
\end{figure*}

\begin{table}[htpb]
\centering
\renewcommand{\arraystretch}{1.4}
\caption{The influence of $\lambda$ in MAFNet.}
\begin{tabular}{c|ccccc}
\toprule
~~~ $\lambda$ ~~~ & 0.000 & 0.005 & 0.010 & 0.015 & 0.020 \\
\midrule
    PSNR $\uparrow$ & 33.49 & 33.74 & 34.03 & 33.80 & 33.78 \\ 
\bottomrule
\end{tabular}
\label{lambda fig}
\end{table}

\subsection{Experiments on Remote-Sensing HSI Datasets}

To verify the robustness and denoising performance of the proposed MAFNet, we conduct extensive experiments on three remote-sensing HSI datasets. The first dataset is the  \emph{Pavia University dataset}. It contains 103 bands, and the spatial size of the image is  $610\times340$ pixels. It was captured by the reflective optics system imaging spectrometer sensor (ROSIS-3) over the Pavia University, Italy. The second dataset is the \emph{Urban dataset}. It was captured by the HYDICE sensor. The sensor provides 210 bands ranging from 400 nm  to 2500 nm. In order to further verify the denoising performance of MAFNet on real hyperspectral noise, we introduce the Indian Pines dataset, which is captured by the AVIRIS sensor and contains 220 bands with a resolution size of $145\times145$ pixels.

\emph{1) Results on the Pavia University Dataset with Mixture Noise.} We added mixture noise on the Pavia University dataset, and the experimental results are listed in Table \ref{pavia table}. It can be easily seen that the proposed model achieves the highest quantitative metrics. The corresponding visual results are provided in Fig. \ref{pavia fig}, and the values of the PSNR and SSIM within different bands of the restored HSI on the dataset are depicted in Fig. \ref{value}. Our method not only effectively removes the complex noise, but also simultaneously preserves the high-frequency texture details. Furthermore, the mean normalized digital number curves by different methods are shown in Fig. \ref{rdn}, which demonstrates that the proposed MAFNet effectively removes the complex noise without introducing obvious spectral distortion.

As demonstrated in Fig. \ref{value}, the proposed method achieves the best PSNR values on almost all spectral bands. We find that nearly all the methods perform differently across spectral bands. It is mainly caused by the denoising difficulties among different spectral bands, since the intrinsic noise levels from different spectral bands are different. In addition, due to the characteristics of the hyperspectral sensor, the spatial details of the different spectral bands are different. Hence, nearly all the methods perform differently across spectral bands.

\emph{2) Results on the Urban Dataset with Mixture Noise.} To further validate the efficacy of MAFNet in denoising, we added severely polluted noise to the Urban dataset. Figs. \ref{urban fig} and \ref{urban_2 fig} demonstrate that deep learning-based methods are effective in removing severely polluted noise. Notably, the proposed MAFNet achieves the best denoising performance among all the methods.

\emph{3) Results on the Indian Pines Dataset with Real-World Noise.} Some bands in the Indian Pines dataset are seriously corrupted by the atmosphere and water, and are polluted by complex noises. We show the denoising result of various methods on this dataset in Fig. \ref{india fig}. It can be observed that MAFNet can still achieve satisfactory denoising performance in real hyperspectral noise. Furthermore, it performs better in preserving spatial details.

\renewcommand{\arraystretch}{1.5}
\begin{table}[bpt]
\caption{Ablation study of co-attention and AIN .}
\centering
\begin{tabular}{c|ccc} 
    \toprule
	~~~ Method ~~~ & Co-attention & ~~~ AIN ~~~  &  ~~~ PSNR $\uparrow$ ~~~ \\ 
	\midrule
	(a) & \checkmark &  & 33.93 \\
	(b) & & \checkmark &  33.85\\
	(c) & \checkmark & \checkmark & 34.03 \\
	\bottomrule
\end{tabular}
\label{table_attn_ain}
\end{table}

\subsection{Ablation Study and Parameter Sensitivity Analysis}

\textbf{Co-attention and AIN.} The critical parts of the proposed method are the co-attention and AIN modules. In order to validate the effectiveness of both modules, we comprehensively compare three variants on the Pavia University dataset. (a) Basic network with co-attention module. (b) Basic network with AIN module. (c) Basic network with both co-attention and AIN module (our proposed MAFNet). The experimental results on the Pavia University dataset are given in Table \ref{table_attn_ain}. It is evident that the combination of co-attention and AIN achieves the best denoising performance. 

\begin{figure}[htb]
\centering
\includegraphics [width=3.0in]{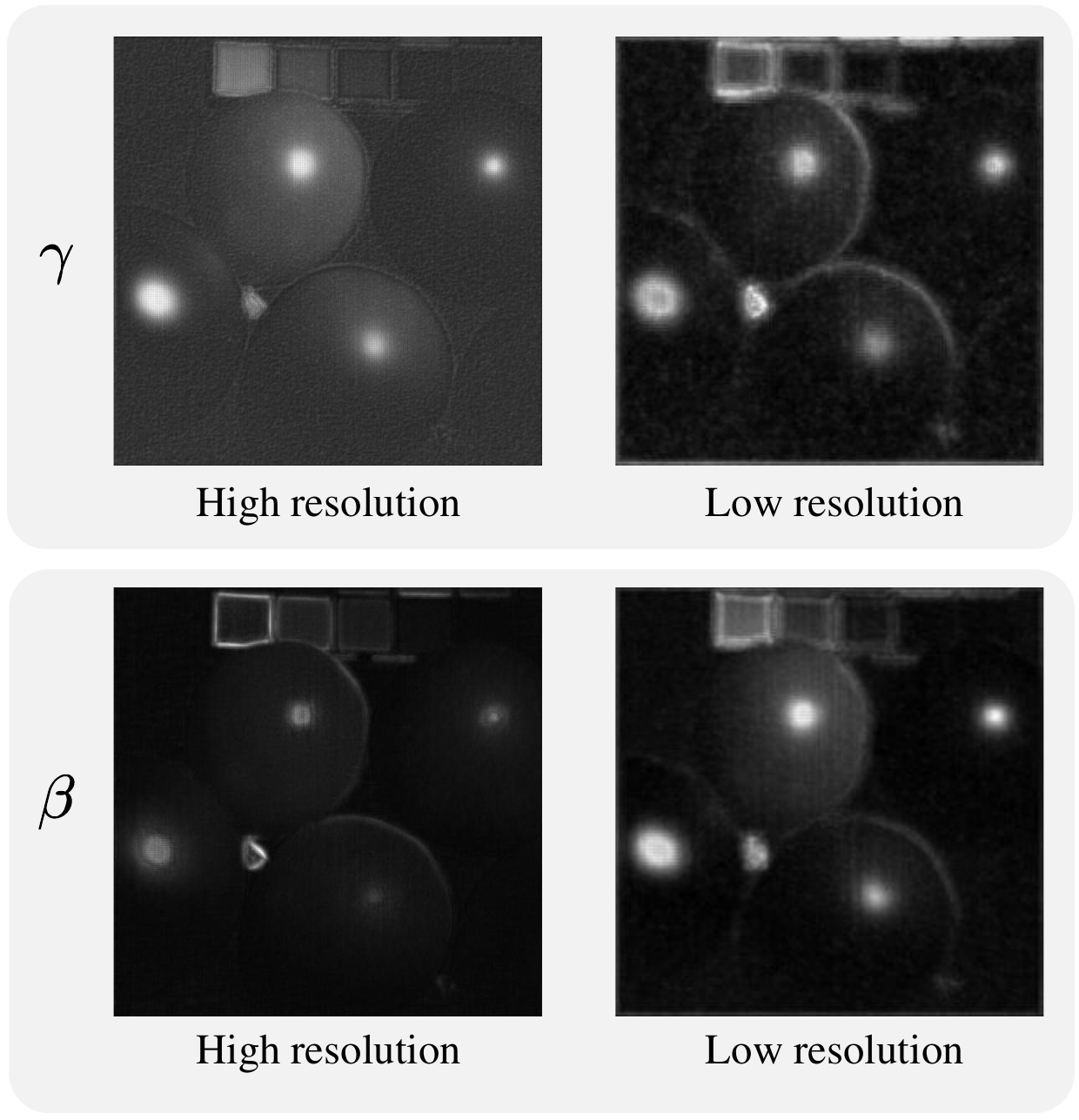}
\caption{Visualization of $\gamma$ and $\beta$ in AIN module.}
\label{fig_Visual_AIN}
\end{figure}

The AIN module provides semantic guidance via channel-wise normalization and pixel-wise affine transformation. The transform parameters $\gamma$ and $\beta$ are visualized in Fig. \ref{fig_Visual_AIN}. In the AIN module, $\gamma$ calibrates the input feature and highlights the important regions. As can be observed that in the low-resolution branch, $\gamma$ highlights the object boundary and thus transfers the basic structure to the high-resolution branch. $\beta$ is used as the complementary information for feature calibration in the AIN module, providing more details to complete the denoising task.

\renewcommand{\arraystretch}{1.5}
\begin{table}[htb]
\caption{Different feature fusion schemes for co-attention module.}
\centering
\begin{tabular}{c|ccc} 
    \toprule
	~~~ Setting ~~~ & ~~ Concat ~~ & ~ Multiply ~  &  ~~~ Ours ~~~ \\ 
	\midrule
	PSNR $\uparrow$ & 33.96 & 33.93 & 34.03 \\
	\bottomrule
\end{tabular}
\label{table_fusion}
\end{table}

\textbf{Feature Fusion in Co-Attention Module.} Here, we discuss the multiscale feature fusion in co-attention module. In the proposed MAFNet, multiscale features are fused by element-wise summation. We also designed two other schemes, and the experimental results on the Pavia University dataset are shown in Table \ref{table_fusion}. The ``Concat" uses concatenation for multiscale feature fusion, and employs $1\times1$ convolution layer to reduce the channels. The ``Multiply" uses element-wise multiplication for feature fusion. Compared with ``Concat" and ``Multiply", the proposed method could achieve slightly better denoising performance.

\renewcommand{\arraystretch}{1.5}
\begin{table}[htb]
\caption{Different attention schemes for co-attention module.}
\centering
\begin{tabular}{c|ccc} 
    \toprule
	~~~ Setting ~~~ & ~~ Split ~~ & ~ C-Attn ~  &  ~~~ Ours ~~~ \\ 
	\midrule
	PSNR $\uparrow$ & 33.90 & 39.87 & 34.03 \\
	\bottomrule
\end{tabular}
\label{table_attention}
\end{table}

\textbf{Split Attention and Self-Calibration.} 
Two attention mechanisms are employed in the co-attention module: Split attention and self-calibration. To verify the effectiveness of both mechanisms, we design three variants, as shown in Table \ref{table_attention}. The ``Split" only uses the concatenation and split part in co-attention module, and the self-calibration is removed. The ``C-Attn" uses the channel attention \cite{hu2018squeeze} instead of the concatenation and split part in co-attention module, and the self-calibration is removed. In the proposed MAFNet, both split attention and self-calibration are used. The experimental results on the Pavia University dataset are shown in Table \ref{table_attention}. It can be observed that ``Split" slightly outperforms ``C-Attn", since multiscale features are adaptively emphasized by split attention. The proposed MAFNet performs the best by combining split attention and self-calibration.

\textbf{Global Gradient Regularizer.}
In our designed loss function, $\lambda$ is a critical parameter that affects the global gradient regularizer, and can affect the denoising performance. We evaluate the denoising performance by taking different $\lambda$ on the ICVL dataset while keeping the network unchanged. The results are shown in Table \ref{lambda fig}, and it can be observed that the proposed MFANet reaches the bet PSNR value when $\lambda=0.010$. Therefore, in our implementation, $\lambda$ is set to 0.010.

\textbf{Number of Channels.}
We explore the influence of the channel number on the denoising performance, and design three variants of MAFNet according to the channel number and model size, i.e., MAFNet-S, MAFNet-B, and MAFNet-L. The channel numbers of three scales of MAFNet-S are set as (32, 64, and 128). Then, the corresponding channel numbers in MAFNet-B and MAFNet-L are set as (64, 128, and 256) and (128, 256, and 512), respectively. The denoising performance of three variants and other deep learning methods are illustrated in Table \ref{time_parameters}. It can be observed that MAFNet-L achieves the best PSNR value, but its computational complexity is large. MAFNet-S achieves good denoising performance while it is rather computationally efficient. It should be noted that MAFNet-B achieves excellent denoising performance while its computational complexity is within an acceptable range. Therefore, in our implementations, we set the channel numbers of three scales as (64, 128, and 256). Compared with other deep learning-based methods, our method exhibits impressive performance in FLOPs and the number of parameters. Specifically, although the FLOPs of our method are higher compared than HSID-CNN, the denoising performance of our method is much better. Compared with MemNet and QRNN3D, the proposed method is more computationally efficient. 

\begin{table}[htpb]
\centering
\renewcommand{\arraystretch}{1.4}
\caption{Denoising performance and the number of parameters comparison.}
\begin{tabular}{c|ccc}
\toprule
~~~ Method ~~~ & PSNR $\uparrow$ & FLOPs (G) $\downarrow$ & Params (M) $\downarrow$\\
\midrule
    MemNet \cite{tai17iccv} & 29.27 & 31.61 & 1.92 \\ 
    HSID-CNN \cite{yuan19tgrs} & 30.38 & 6.61 & 0.63 \\ 
    QRNN3D \cite{wei21tnnls} & 33.83 & 47.62 & 3.37 \\
    MAFNet-S  & 32.87 & 5.44 & 2.03 \\  
    MAFNet-B  & 34.03 & 20.15 & 7.55 \\ 
    MAFNet-L  & 34.21 & 77.71 & 29.21 \\    
\bottomrule
\end{tabular}
\label{time_parameters}
\end{table}

\section{Conclusion and Future Work}

In this paper, we present a multiscale adaptive fusion network for hyperspectral image noise reduction. Two key components contribute to improving the hyperspectral image denoising: A progressively multiscale information aggregation framework and co-attention fusion module. Specifically, a set of multiscale images are generated and fed into a coarse-fusion network to exploit the contextual texture correlation. Thereafter, a fine fusion network is followed to exchange the information across the parallel multiscale subnetworks. Ultimately, the co-attention fusion module adaptively emphasizes informative features from different scales and reinforces the discriminative learning capability for denoising. Experiments on both synthetic and real HSI datasets verified the superiority of the proposed method compared with other state-of-the-art HSI denoising methods. 

Although the MAFNet in this paper exhibits outstanding denoising performance, the utilization of multi-scale branches for feature extraction also results in an increase in the number of parameters and computational complexity. Consequently, in the future, we will concentrate on devising lightweight HSI denoising techniques.

\bibliography{source}
\bibliographystyle{IEEEtran}

\begin{IEEEbiography}[{\includegraphics[width=1in,height=1.25in,clip,keepaspectratio]{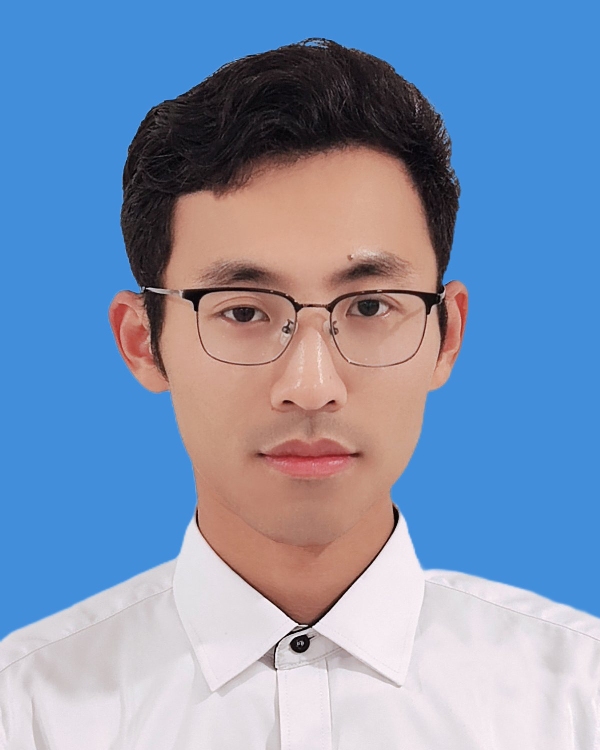}}]{Haodong Pan} 
received the B.Eng. degree in computer science and technology from the Ocean University of China, Qingdao, China, in 2020. He is currently pursuing the M.Sc. degree
in computer science with the School of Information Science and Technology, Ocean University of China, Qingdao, China.

His current research interests include computer
vision and remote sensing image processing.

\end{IEEEbiography}

\begin{IEEEbiography}[{\includegraphics[width=1in,height=1.25in,clip,keepaspectratio]{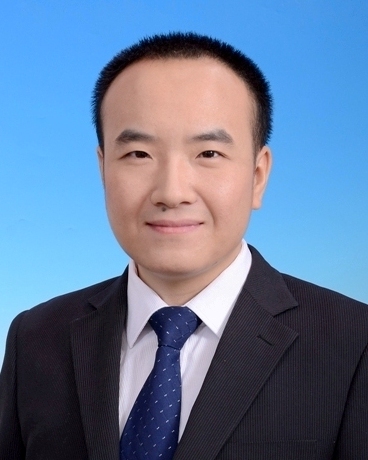}}]{Feng Gao} (Member, IEEE)
received the B.Sc degree in software engineering from Chongqing University, Chongqing, China, in 2008, and the Ph.D. degree in computer science and technology from Beihang University, Beijing, China, in 2015.

He is currently an Associate Professor with the School of Information Science and Engineering, Ocean University of China. His research interests include remote sensing image analysis, pattern recognition and machine learning.
\end{IEEEbiography}

\begin{IEEEbiography}[{\includegraphics[width=1in,height=1.25in,clip,keepaspectratio]{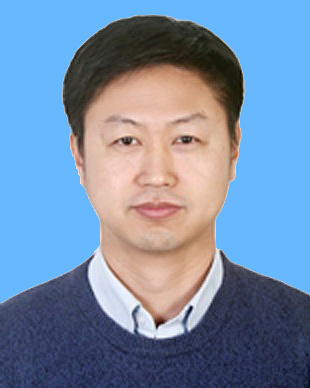}}]{Junyu Dong}
 (Member, IEEE) received the B.Sc. and M.Sc. degrees from the Department of Applied Mathematics, Ocean University of China, Qingdao, China, in 1993 and 1999, respectively, and the Ph.D. degree in image processing from the Department of Computer Science, Heriot-Watt University, Edinburgh, United Kingdom, in 2003.

He is currently a Professor and Dean with the School of Computer Science and Technology, Ocean University of China. His research interests include visual information analysis and understanding, machine learning and underwater image processing.
\end{IEEEbiography}

\begin{IEEEbiography}[{\includegraphics[width=1in,height=1.25in,clip,keepaspectratio]{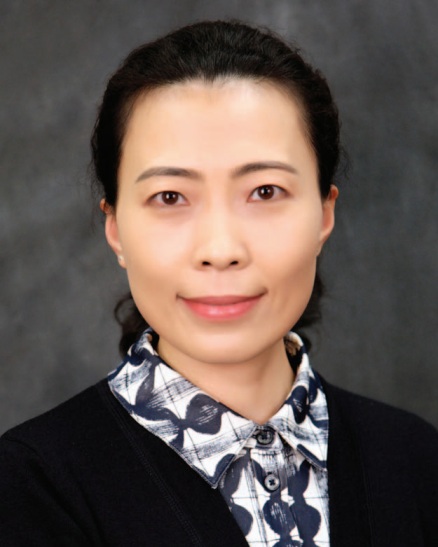}}]{Qian Du}
(Fellow, IEEE) received the Ph.D. degree in electrical engineering from the University of Maryland at Baltimore, Baltimore, MD, USA, in 2000.

She is currently the Bobby Shackouls Professor with the Department of Electrical and Computer Engineering, Mississippi State University, Starkville, MS, USA. Her research interests include hyperspectral remote sensing image analysis and applications, and machine learning.

Dr. Du was the recipient of the 2010 Best Reviewer Award from the IEEE Geoscience and Remote Sensing Society (GRSS). She was a Co-Chair for the Data Fusion Technical Committee of the IEEE GRSS from 2009 to 2013, the Chair for the Remote Sensing and Mapping Technical Committee of International Association for Pattern Recognition from 2010 to 2014, and the General Chair for the Fourth IEEE GRSS Workshop on Hyperspectral Image and Signal Processing: Evolution in Remote Sensing held at Shanghai, China, in 2012. She was an Associate Editor
for the \textsc{Pattern Recognition}, and \textsc{IEEE Transactions on Geoscience and Remote Sensing}. From
2016 to 2020, she was the Editor-in-Chief of the \textsc{IEEE Journal of Selected Topics in Applied Earth Observation and Remote Sensing}. She is currently a member of the IEEE Periodicals Review and Advisory Committee and SPIE Publications Committee. She is a Fellow of SPIE-International Society for Optics and Photonics (SPIE).

\end{IEEEbiography}

\end{document}